\documentclass[12pt]{article}
\usepackage[T1]{fontenc}
\usepackage[utf8]{inputenc}
\usepackage[english]{babel}
\usepackage[a4paper,top=3.5 cm,bottom=2.5 cm,left=2cm,right=2cm,heightrounded,bindingoffset=5mm]{geometry}
\usepackage{authblk}
\usepackage[authoryear]{natbib}
\usepackage{amsmath}
\usepackage{amsfonts}
\usepackage{amssymb}
\usepackage{graphicx}
\usepackage{latexsym}
\usepackage{array}
\usepackage{booktabs}
\usepackage{float}
\usepackage{multirow}
\usepackage{color}
\usepackage[dvipsnames]{xcolor}
\usepackage{algorithm} 
\usepackage{soul}
\usepackage{rotating}
\floatname{algorithm}{Algorithm} 
\renewcommand{\thealgorithm}{\unskip} 
\usepackage{setspace}
\usepackage{marvosym}

\date{}

\begin{document}
\author[1]{Fabio Divino 
	\footnote{\Letter \; \textit{fabio.divino@unimol.it}}
}
\author[2]{Denekew Bitew Belay}
\author[3]{Nico Keilman}
\author[4]{Arnoldo Frigessi}

\affil[1]{\textit{Division of Physics, Computer Science and Mathematics - DiBT, University of Molise}}
\affil[2]{\textit{School of Mathematical and Statistical Sciences, Hawassa University}}
\affil[3]{\textit{Department of Economics, University of Oslo}}
\affil[4]{\textit{Oslo Center for Biostatistics and Epidemiology, University of Oslo and Oslo University Hospital}}

\title{Bayesian Modelling of Lexis Mortality Data}

\maketitle

\begin{abstract}
In this work we present a spatial approach to model and investigate mortality data referenced over a Lexis structure. We decompose the force of mortality into two interpretable components: a Markov random field, smooth with respect to time, age and cohort which explains the main pattern of mortality; and a secondary component of independent shocks, accounting for additional non-smooth mortality. Inference is based on a hierarchical Bayesian approach with Markov chain Monte Carlo computations. We present an extensive application to data from the Human Mortality Database about 37 countries. For each country the primary smooth surface and the secondary surface of additional mortality are estimated. The importance of each component is evaluated by the estimated value of the respective precision parameter. For several countries we discovered a band of extra mortality in the secondary surface across the time domain, in the age interval between 60 and 90 years, with a slightly positive slope. The band is significant in the most populated countries, but might be present also in the others. The band represents a significant amount of extra mortality for the elderly population, which is otherwise incompatible with a regular and smooth dynamics in age, year and cohort.
\end{abstract}


\newpage
\section{Introduction}\label{intro}
In demography and epidemiology, the analysis of mortality data referenced over a Lexis structure allows to capture the simultaneous effects of age, period, and cohort (with their potential interactions); extract relevant mortality patterns; detect features of the population dynamics; and predict mortality trends by different ages and causes of death. The notion of Lexis surface was introduced by Arthur and Vaupel \citep{artvau:1984} to generalize the concept of Lexis diagram: population data indexed over small time and age intervals can be interpreted as a surface with constant values in each time-age square. When the time and the age intervals are short, the discrete surface can be approximated by a continuous surface which may be interpreted as a density across the time and age domains \citep{artvau:1984}. One of the first applications can be found in \cite{casalt:1985}, in which the authors introduce the contour map technique to represent surfaces of mortality rates.\\
At first, the approach was descriptive and oriented to develop statistical methods and graphical tools to display the data, useful in comparative studies. For instance in \cite{casalt:1987}, the authors investigate differences and similarities between the mortality patterns observed in the surfaces of France and Italy, in the period 1900-1979. Since the publication of the seminal paper by \cite{leecar:1992}, interest has been oriented towards the development of statistical models and computational methods to model mortality data in terms of surfaces. The main reason lies in the potential use of Lexis mortality data to forecast mortality dynamics and predict population scenarios by different ages. These aspects are particularly relevant for policy making in social security and public health, and also for risk evaluations by insurance companies. The Lee-Carter (LC) model \citep{leecar:1992} represents the reference model in the literature and it is currently adopted as benchmark by several international institutions including the US Census Bureau and the United Nations. The LC model is a parametric model that can be used to produce mortality and life expectancy forecasts through a two steps procedure: the estimation of parameters by the ordinary least squares method with singular value decomposition and the prediction of mortality rates referring to age and time patterns by the autoregressive integrated moving average method.\\ 
Extensions of the LC approach have been proposed by several authors: Renshaw and Haberman modify the LC model to account for mortality reduction factors \citep{renhaa:2003} and propose a generalized linear regression approach by modelling the data with Poisson distributions \citep{renhab:2003}. In \cite{brouhn:2002} a Poisson log-bilinear regression model is proposed to improve the flexibility of the LC approach while \cite{lilee:2005} use the LC method to forecast the mortality of a group of populations. Further extensions of the LC model can be found in \cite{lee:2000}, \cite{leemil:2001}, and \cite{lialt:2009}, while the use of the LC approach in a Bayesian framework is presented in \cite{czado:2005}, \cite{pedroz:2006}, \cite{katrie:2015}, and \cite{wisnal:2015}.\\ 
A Bayesian model, different from the LC approach, is presented in \cite{dellap:2001} to extend the parametric model introduced by \cite{helpol:1980} and account for incomplete life tables. With a different perspective \cite{cairns:2011} present a Bayesian approach to model the joint behaviour in mortality patterns of two dependent populations.\\
Another approach to the modelling and the estimation of mortality surfaces considers the use of nonparametric smoothing techniques. For instance, in \cite{currie:2004} the application of the P-spline method is presented, in \cite{luoma:2012} smoothing splines are considered in a Bayesian setting, while \cite{lialt:2016} implement a two dimensional kernel smoother to account for cohort effects. An application of a semiparametric approach with the use of the B-spline technique can be found in \cite{barbi:2011} in which the relative importance on the elderly mortality of the cohort effect versus the period effect is discussed.\\ 
The problem of the forward mortality surface, that is the prediction of the whole period-age structure of mortality, is addressed in \cite{bauer:2012} by the use of a finite dimensional Brownian motion while \cite{gaosha:2017} present a functional data analysis to forecast multiple populations. Comparative studies of different approaches and models are reported in \cite{habren:2011}, \cite{giacom:2012}, \cite{danesi:2015}, and \cite{novokr:2016}.\\
\cite{carste:2007} presents and discusses period-age-cohort models with disease data. The role of period-age, age-cohort and period-age-cohort models in epidemiology is reviewed in \cite{clayta:1987, claytb:1987}, particularly with respect to the analysis of cancer data.\\
A fundamental reference for the interpretation of mortality surfaces is the seminal paper by \cite{vaupel:1979} in which the notion of frailty is introduced to interpret the extra variation observed in mortality rates.\\
In the present work we propose a new spatial statistics approach to estimate the force of mortality across a Lexis structure in order to decompose the data into two interpretable components: a primary component accounting for the main smooth effect of period, age, and cohort; and a secondary component representing additional mortality in excess or in defect of that primary pattern. We present an extensive analysis using data from the Human Mortality Database about 37 countries. For several countries we observed a band of extra mortality in the secondary surface across the time domain, in the age interval between 60 and 90 years, with a slightly positive slope. The band is observed in the most populated countries and represents an overdispersion effect due to the presence of heterogeneity in the data.\\
The paper is organized as follows. Section \ref{s2} reviews the methodological background concerning the analysis of mortality data across a Lexis structure, introduces the basic idea of our approach, and discusses the similarities with other models. In Section \ref{s3} we present the hierarchical Bayesian model with mathematical details: likelihood, priors, and posterior. Furthermore, the Bayesian estimation of the two components and computational aspects of the Markov Chain Monte Carlo (MCMC) algorithm are discussed in detail. Section \ref{s4} describes the datasets we considered whilst Section \ref{s5} presents results and comments. In particular, estimates of the Bayesian mortality surface with the primary and secondary components are presented for the cases of Sweden and Italy, countries with different demographic transitions and mortality patterns. Finally, in Section \ref{s6} conclusions are drawn and future developments are briefly discussed.

\section{A spatial approach for Lexis mortality data}\label{s2}
A Lexis diagram is a two dimensional system of coordinates, for example representing calender time and age, fundamental to investigate demographic phenomena such as the mortality events occurring in populations of interest \citep{vandes:2001}. A Lexis diagram for individuals in a population allows to plot the death and life of each individual. Each death event is represented by a point, with continuous time and age coordinates, whilst each individual of the population is represented by a segment with unit slope and connecting the moment of birth on the time coordinate with the point event.  In the case of mortality events, the diagonal segments represent the individual life lines and the deaths of a population can be interpreted as the realization of a point process on the time-age coordinates system \citep{keidin:1990}.\\ 
The fundamental relevance of the Lexis diagram consists in the possibility to jointly represent three demographic dimensions: period, age, and cohort. In fact, across the diagonal directions with unit slope, it is possible to interpret the events as cohort related.\\
In this paper, we consider a new approach of the period-age-cohort structure and it is useful to interpret the diagram as a graph. We consider a finite time domain $\mathcal{T}=\{1,...,T\}$ (e.g. a set of calendar years), a finite age domain $\mathcal{A}=\{1,...,A\}$ (e.g. a set of age classes), and the two dimensional Lexis lattice $\mathcal{L}=\mathcal{T}\times\mathcal{A}$, with $N=T \times A$ knots. For each knot $(t, j) \in \mathcal{L}$, we denote by $y_{t j}$ the observed death count, by $n_{t j}$ the exposed population, and by $\mu_{t j}$ the force or intensity of mortality at the time $t$ and the age $j$.\\
Under the assumption that each death count $y_{tj}$ is the realization of an independent Poisson random variable with respective mean $\lambda_{tj}=\mu_{tj}n_{tj}$, the maximum likelihood estimate of the mortality intensity $\mu_{t j}$ can be obtained in each knot of the Lexis graph as the empirical mortality rate 
$$m_{t j}=\dfrac{y_{t j}}{n_{t j}}.$$ 
When mortality is represented in log-scale on the Lexis graph, the empirical rates produce an empirical mortality surface \citep{artvau:1984, vaupel:1997} that shows the main pattern of the force of mortality. On the other hand, if we are also interested to investigate the simultaneous effects of period, age, and cohort or the presence of latent structures due to heterogeneity factors, a statistical modelling approach is necessary \citep{carste:2007}.\\
In 1992, Lee and Carter introduced their seminal model \citep{leecar:1992}. At each point $(t, j)$ of the Lexis graph, the mortality rate is modelled on the log-scale as the additive result of a baseline schedule across the age domain, $a(j)$, and a period-age interaction, defined by a multiplicative term, $r(t,j)=k(t)b(j)$, that is 
\begin{equation} \label{leecarter} \nonumber
\log (m_{tj})=a(j)+k(t)b(j)+ \epsilon_{tj},
\end{equation} 
where $\epsilon_{tj}$ is an independent Gaussian error.
The LC model captures the main pattern of the underlying force of mortality and the interaction period-age but it does not include a direct cohort component, accounting for the effect of potential selection factors \citep{renhab:2006}. Furthermore, some relevant issues arise on the identification of the parameters and the computation of their estimates so that constraints on the parameter estimates of $k(t)$ and $b(x)$ are necessary  \citep{renhaa:2003}.\\ 
In the literature several extensions of the LC model have been considered, particularly with different representations of the interaction term $r(t,j)$ and with different assumptions about the sampling error \citep{brouhn:2002,renhaa:2003,renhab:2006}. Nevertheless, the common approach consists in modelling the components $a(j)$ and $r(t,j)$ in terms of factor effects or regression functions.\\
Our proposal originates from a different point of view. It is very reasonable to assume that the force of mortality at any point $(t, j)$ of the Lexis lattice $\mathcal{L}$ is comparable and rather similar to the force of mortality of knots close in time and age, which we call here neighbouring knots.
Therefore the mortality pattern should be rather smooth with respect to time, age, and cohort.\\
The smoothness of functions defined on a lattice system is often an important assumption in order to derive powerful and coherent statistical  models. For instance, in the field of geographical epidemiology, the general risk of a certain disease is supposed to be smooth across the spatial domain of interest \citep{lawson:2006}; in image analysis, the intensity measured at some pixel of a satellite or medical image can be typically assumed similar to the measurements recorded in its nearby pixels \citep{listan:2009, winkle:2003}. Spatial statistics is a collection of models and inferential methods to study functions which change smoothly in the space where they are defined \citep{banalt:2004, cressi:1991, schgot:2005} and we follow this approach in order to model mortality data across the Lexis lattice. In fact, by analogy, a mortality surface can be considered as an image in which the domain of the pixels is defined across time and age instead of latitude and longitude.\\ 
In each knot $(t, j) \in \mathcal{L}$ we assume that the mortality intensity results from the effect of two components, $x_{tj}$ and $z_{tj}$, acting at different regularity scales of the Lexis structure. The first component $x_{tj}$ is a locally smooth term which takes into account for the main effect of time, age, and cohort. In addition to this smooth component, we account for additional mortality in excess or in defect. This second feature, that could be the result of heterogeneous latent factors acting on the population of interest, is modelled by a set of independent shocks $z_{tj}$.\\ 
Through the canonical link of the Poisson likelihood, we assume the following log-linear model
\begin{equation} \label{log1}
\log (\mu_{tj})=c_0+x_{tj}+z_{tj},
\end{equation}
in which $c_0$ is a fixed offset as, for instance, the general baseline risk of the whole surface at the log-scale. We assume that the two components in Equation \eqref{log1} are random effects: the first one, $x_{tj}$, is locally structured and similar to the effects in the neighbouring knots while the second one, $z_{tj}$, does not have any regular structure on the Lexis lattice. Notice that this second component is not a residual component but a term accounting for additional mortality with a non smooth pattern. We need to specify the definition of similarity and nighbourhood and details are given in Section \ref{s3}.\\
The idea to represent the mortality intensity by two components is not new. It was introduced by Vaupel, Manton, and Stallard in their seminal paper \citep{vaupel:1979} to account for the potential heterogeneity introduced by latent selection factors, the frailty by cohort. The authors considered the possibility that individuals, even though in the same cohort, can experience the event of interest (death) with different attitude or susceptibility: the individual frailty. Then, under the assumption that two mortality intensities at the individual level are proportional to each other by the ratio of the respective frailties, for each point $(t,j) \in \mathcal{L}$, the authors derived the following relation
\begin{equation} \label{vaupel}
\mu_{tj}=\mu_{tj,1}\zeta_{tj},
\end{equation}
in which $\mu_{tj,1}$ is the force of mortality for a standard individual with unit frailty while $\zeta_{tj}$ is the expected frailty at the population level, obtained by averaging the individual frailty through a Gamma distribution, see \cite{vaupel:1979} for details. At the log-scale, Equation \eqref{vaupel} is similar to Equation \eqref{log1}, but we make different assumptions on the two components which allow us to discover new features in the Lexis surface.

\section{A hierarchical Bayesian model for the force of mortality}\label{s3}
We adopt a Bayesian approach and formalize a hierarchical model for Equation \eqref{log1} in which the randomness of the two components, the locally structured effect $x_{tj}$ and the independent shock $z_{tj}$, is modelled in terms of their prior probability distributions. These priors, and the additional hyperpriors, are combined with the Poisson likelihood through the Bayes theorem and in order to derive the joint posterior distribution, necessary to make inference on the quantities of interest. A similar approach was introduced in 1991 by Besag, York and Mollie \citep{besalt:1991} in the field of disease mapping and then widely applied, including ecological analysis \citep{clayal:1993, ricbes:2003} and geographical demography \citep{divino:2009}.

\subsection{Likelihood}
Given the Lexis lattice $\mathcal{L}$, for each knot $(t,j) \in \mathcal{L}$, we assume that the mortality count $y_{tj}$ is the realization of a conditional independent Poisson random variable given the mean $\lambda_{tj}=\mu_{tj}n_{tj}$. Therefore, the full likelihood is given by
$$
L(\mu;\mathbf{y})=\prod_{t \in \mathcal{T}}\prod_{j \in \mathcal{A}} \dfrac{(n_{tj}\mu_{tj})^{y_{tj}}e^{-n_{tj}\mu_{tj}}}{y_{tj}!},
$$
in which $\mu$ and $\mathbf{y}$ denote the sets $\{ \mu_{tj}: t \in \mathcal{T}, j \in \mathcal{A} \}$ and $\{ y_{tj}: t \in \mathcal{T}, j \in \mathcal{A} \}$ respectively.
Furthermore, we assume that the Poisson mean $\lambda_{tj}$, at any point $(t,j) \in \mathcal{L}$, transformed by the canonical logarithmic link, can be represented as follows
\begin{equation} \label{log2}
\log (\lambda_{tj})=\log (E_{tj}) + x_{tj} +z_{tj},
\end{equation}
where $E_{tj}$ represents the expected mortality count under the constant baseline intensity $\mu_0$ acting on the whole surface, that is $E_{tj}=\mu_0 n_{tj}$. It means that the expected count $\lambda_{tj}$, at any knot $(t,j) \in \mathcal{L}$, is equal to the expected count under a constant intensity, $E_{tj}$, adjusted by the sum (at the exponential scale) of the two effects $x_{tj}$ and $z_{tj}$, that is $\lambda_{tj}=E_{tj}e^{x_{tj}+z_{tj}}$. Furthermore, from Equation \eqref{log2} it follows 
\begin{equation} \label{log3}
\log (\mu_{tj})=\log (\mu_0) + x_{tj} +z_{tj},
\end{equation}
that models the force of mortality $\mu_{tj}$ in terms of $\mu_0$, $x_{tj}$, and $z_{tj}$. Notice that Equation \eqref{log3} corresponds to Equation \eqref{log1} with $c_0=\log(\mu_0)$. The two terms $x_{tj}$ and $z_{tj}$ represent different features of the mortality intensity: $x_{tj}$ is the locally structured component accounting for the smooth effect of time, age, and cohort whilst $z_{tj}$ is the no structured component accounting for additional (but not residual) mortality in excess or in defect. Next, we define $x_{tj}$ and $z_{tj}$ precisely.

\subsection{Priors and hyperpriors} 
\begin{figure}
\centering
\includegraphics[scale=0.5]{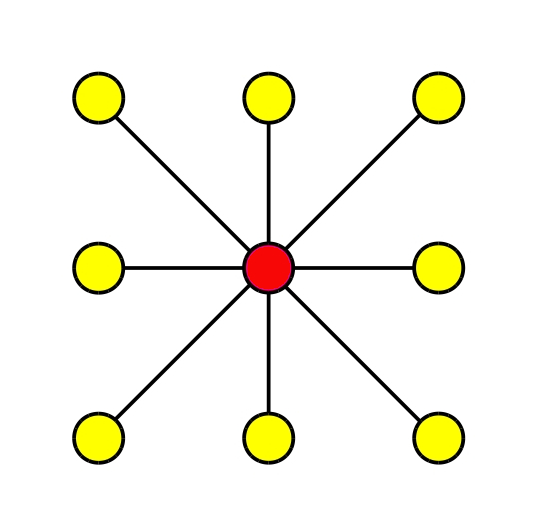}
\caption{The first order Markov neighbourhood system on the Lexis graph.}\label{fig:graph}
\end{figure}

Let us denote by $\mathbf{x}$ and $\mathbf{z}$ the sets of the components $x_{tj}$ and $z_{tj}$ respectively, that is $\mathbf{x}=\{ x_{tj}: t \in \mathcal{T}, j \in \mathcal{A} \}$ and $\mathbf{z}=\{ z_{tj}: t \in \mathcal{T}, j \in \mathcal{A} \}$. We assume that the smooth term $\mathbf{x}$ is apriori distributed as an intrinsic Gaussian Markov random field \citep{besag:1974, cressi:1991, listan:2009, ruehel:2005} with pairwise interactions
\begin{equation} \label{gmrf} \nonumber
p(\mathbf{x} |\gamma_x) \propto ( \gamma_x)^{\frac{n}{2}} \exp \left\lbrace  -\frac{\gamma_x}{2} \sum_{(t,j) \sim (s,i)}  (x_{tj}-x_{si})^2 \right\rbrace ,
\end{equation}
in which the symbol $ \sim $ denotes the symmetric relation of the first order Markov neighbourhood \citep{listan:2009} among the sites of $\mathcal{L}$, and $\gamma_x$ is the positive precision parameter of the field $\mathbf{x}$, governing the strength of these pairwise interactions. We consider two knots $(t,j)$ and $(s,i)$ as neighbours if they are adjacent along the three direction of the Lexis graph: time (latitudinal), age (longitudinal), and cohort (diagonal); and along the diagonal with negative unit slope to account for further correlation between parallel cohorts. Therefore, the first order Markov neighbourhood system of a specific knot in red on the Lexis lattice in Figure \ref{fig:graph} is represented by the eight adjacent knots, in yellow in Figure \ref{fig:graph}. In other words, in each point $(t,j)$ of the Lexis graph, given the neighbours, the component $x_{tj}$ represents a conditional autoregressive term with Gaussian distribution \citep{cressi:1991}.
\begin{center}
	< Figure \ref{fig:graph} >
\end{center}
As $\mathbf{z}$ is a set of independent variables, we assume that each $z_{tj}$ is a priori distributed as a Gaussian random variable centered in zero and with positive precision parameter $\gamma_z$, so that the joint distribution of $\mathbf{z}$ is given by
\begin{equation} \label{mvn} \nonumber
p(\mathbf{z} |\gamma_z) \propto ( \gamma_z)^{\frac{n}{2}} \exp \left\lbrace  -\frac{\gamma_z}{2} \sum_{(t,j)} z_{tj}^2 \right\rbrace .
\end{equation}
On top of these two priors we need to define the hyperpriors for the precision parameters. Following \cite{mollie:1999}, we assume that $\gamma_x$ and $\gamma_z$ are apriori independently distributed as Gamma variables, that is
\begin{equation} \label{gammax} \nonumber
p(\gamma_x;\alpha_x,\beta_x)=\dfrac{\beta_x^{\alpha_x}}{\Gamma(\alpha_x)} e^{-\beta_x\gamma_x} \gamma_x^{\alpha_x-1},
\end{equation}
and
\begin{equation} \label{gammaz} \nonumber
p(\gamma_z;\alpha_z,\beta_z)=\dfrac{\beta_z^{\alpha_z}}{\Gamma(\alpha_z)} e^{-\beta_z\gamma_z} \gamma_z^{\alpha_z-1},
\end{equation}
where $(\alpha_x, \beta_x)$ and $(\alpha_z, \beta_z)$ are the respective hyperparameters. In this paper we consider vague hyperpriors \citep{congdo:2014} with $\alpha_x=0.01$, $\beta_x=0.01$, $\alpha_z=0.01$, and $\beta_z=0.01$.\\
The offset $\mu_0$ can be estimated consistently by the maximum likelihood estimate 
\begin{equation} \label{rate}
\hat{\mu}_0=\frac{\sum_{(t,j)} y_{tj}}{\sum_{(t,j)} n_{tj}},
\end{equation}
with the assumption that a constant mortality intensity is acting across the whole surface. It is an appropriately weighted average across time of the crude death rate (CDR), where the CDR in a certain year equals all deaths over all exposures that year. It can also be interpreted as the general effect in a log-linear analysis setting. In this paper we compute it as by Equation \eqref{rate} and then fix it in the analysis. Notice that it could also be considered as a further parameter in the Bayesian modelling although it plays simply the role of scale quantity to stabilize the estimation and the computation of the model, and it does not have any effect on the decomposition of the force of mortality.

\subsection{Posterior distribution} The Bayes theorem gives the joint posterior distribution
\begin{eqnarray} \label{posterior} \nonumber
&&p(\mathbf{x},\mathbf{z},\gamma_x,\gamma_z | \mathbf{y}; \alpha_x,\beta_x,\alpha_z,\beta_z,\mu_0) \propto \exp  \left\lbrace  \sum_{(t,j)}  \left[  y_{tj}(x_{tj}+z_{tj})-\mu_0 n_{tj}e^{x_{tj}+z_{tj}} \right. \right. \\ \nonumber
&&\ \left. - \frac{\gamma_x}{2} \sum_{(s,i) \in \Delta_{tj}} \frac{(x_{tj}-x_{si})^2}{2}  -\frac{\gamma_z}{2} z_{tj}^2  \right]  -\beta_x\gamma_x + \left( \alpha_x-1+\frac{n}{2} \right)\log(\gamma_x) \\ 
&&\left. -\beta_z\gamma_z + \left( \alpha_z-1+\frac{n}{2} \right)\log(\gamma_z)  \right\rbrace,
\end{eqnarray}
in which $\Delta_{tj}$ denotes the set of the sites $(s,i)$ adjacent to $(t,j)$ with respect to the first order Markov neighbourhood system in Figure \ref{fig:graph}. To obtain estimates for $\mathbf{x}$ and $\mathbf{z}$, we have to integrate Equation \eqref{posterior}, but this is not feasible and we need to approximate such integral by Monte Carlo.

\subsection{MCMC computation} To produce samples from the marginal posteriors of $\mathbf{x}$, $\mathbf{z}$, $\gamma_x$, and $\gamma_z$ we implemented an MCMC algorithm with Metropolis steps for the quantities $\mathbf{x}$ and $\mathbf{z}$, and Gibbs Sampler steps for the parameters $\gamma_x$ and $\gamma_z$ \citep{liujun:2004, robcas:2004}.\\
In particular, for each $x_{tj}$ and for each $z_{tj}$, the Metropolis steps use Gaussian proposals centered on the previous value and with variance calibrated to obtain a global acceptance rate approximately between $0.2$ and $0.3$ \citep{robalt:1997, robros:2001}. The Metropolis acceptance probabilities are based on the following local potentials 
\begin{equation}\label{metropolis1}
H_x(x_{tj})=\mu_0n_{tj}e^{x_{tj}+z_{tj}}-y_{tj}x_{tj}+\frac{\gamma_x}{2}\sum_{(s,i) \in \Delta_{tj}} (x_{tj}-x_{si})^2,
\end{equation}
and
\begin{equation}\label{metropolis2}
H_z(z_{tj})=\mu_0n_{tj}e^{x_{tj}+z_{tj}}-y_{tj}z_{tj}+\frac{\gamma_z}{2}z_{tj}^2,
\end{equation}
respectively. Therefore, in the Metropolis step an \textit{old} value is replaced by a \textit{new} proposed value with respective probabilities 
\begin{equation}\label{accept1}
p_x=min\{1,e^{H_x(x_{tj}^{old})-H_x(x_{tj}^{new})}\},
\end{equation} 
and
\begin{equation}\label{accept2}
p_z=min\{1,e^{H_z(z_{tj}^{old})-H_z(z_{tj}^{new})}\},
\end{equation} 
with obvious notation.\\
For $\gamma_x$ and $\gamma_z$, we derive the conditional posteriors as Gamma distributions with conjugate parameters
\begin{equation}\label{parx}
\alpha_x' = \alpha_x+\frac{n}{2} \; \text{and} \; \beta_x'=\beta_x+\frac{1}{2} \sum_{(t,j) \sim (s,i)}  (x_{tj}-x_{si})^2,
\end{equation}
and
\begin{equation}\label{parz}
\alpha_z' = \alpha_z+\frac{n}{2} \; \text{and} \; \beta_z'=\beta_z+\frac{1}{2} \sum_{(t,j)}  z_{tj}^2, 
\end{equation}
respectively, and the Gibbs Sampler steps can be performed directly. Our MCMC algorithm can be summarized in the following scheme.\\
\begin{algorithm}
\caption{Metropolis-Gibbs Sampler}
Initialize $\mathbf{x}$, $\mathbf{z}$, $\gamma_x$, and $\gamma_z$.\\
For each iterations $k$ repeat the following 4 steps.\\
\textbf{Step 1}. For each $(t,j) \in \mathcal{L}$:\\  
(a) sample $x_{tj}^{(k)}$ by Metropolis with local potential \eqref{metropolis1} and acceptance probability \eqref{accept1};\\
(b) sample $z_{tj}^{(k)}$ by Metropolis with local potential \eqref{metropolis2} and acceptance probability \eqref{accept2}.\\
\textbf{Step 2}. Update $(\alpha_x',\beta_x')$ and $(\alpha_z',\beta_z')$ by the rules \eqref{parx} and \eqref{parz} respectively.\\
\textbf{Step 3}. Sample $\gamma_x^{(k)}$ from $Gamma(\alpha_x',\beta_x')$.\\
\textbf{Step 4}. Sample $\gamma_z^{(k)}$ from $Gamma(\alpha_z',\beta_z')$.
\end{algorithm}

\subsection{Bayesian estimates} The MCMC converges to the posterior in Equation \eqref {posterior} when the number of iterations increases to infinity. After the burn-in runs, the MCMC samples are used to compute point estimates for each quantity of interest by the ergodic marginal posterior means
\begin{equation} \label{estimate1} \nonumber
\hat{x}_{tj}=\frac{1}{K}\sum_{k=1}^K x_{tj}^{(k)}, \; \hat{z}_{tj}=\frac{1}{K}\sum_{k=1}^K z_{tj}^{(k)}, \; \hat{\gamma}_{x}=\frac{1}{K}\sum_{k=1}^K \gamma_{x}^{(k)}, \; \text{and} \; \hat{\gamma}_{z}=\frac{1}{K}\sum_{k=1}^K \gamma_{z}^{(k)},
\end{equation}
where $K$ denotes the number of iterations considered after burn-in. Then, a point estimate of the mortality intensity at each knot $(t,j)$ of the Lexis graph $\mathcal{L}$ is obtained by
\begin{equation} \label{estimate2} \nonumber
\hat{\mu}_{tj}=\hat{\mu}_0e^{\hat{x}_{tj}+\hat{z}_{tj}},
\end{equation}
with $\hat{\mu}_0$ as in Equation \eqref{rate}.\\
The estimated smooth component $\hat{x}_{tj}$, for varying $(t,j)$ $\in$ $\mathcal{L}$, allows to interpret the regular mortality pattern that progresses in time, age, and cohort in a locally homogeneous way. Different populations, for example countries, will have different $\hat{\mathbf{x}}=\{\hat{x}_{tj}, (t,j)$ $\in$ $\mathcal{L}\}$ which can be compared to understand population specific smooth mortality structures. The shock component $\hat{z}_{tj}$, on the other hand, collects additional but not residual effects of the mortality surface, which are not smooth. Therefore, $\hat{\mathbf{z}}=\{\hat{z}_{tj}, (t,j)$ $\in$ $\mathcal{L}\}$ allows to determine population specific features that can not be explained by structured smoothness.

\section{The Human Mortality Database}\label{s4}
We considered data at the time September 2016 from the open access Human Mortality Database (HMD, \texttt{www.mortality.org}), a project that since 2002 has involved the Department of Demography at the University of California, Berkeley (US) and the Max Planck Institute for Demographic Research in Rostock (Germany), and supported by other research institutes. When used for international comparisons, HMD data are generally assumed to be of superior quality compared to data from national statistical organizations. The reason is that raw HMD data are processed based on a common state-of-the-art procedure. The adjustments includes handling of death counts with missing ages and estimation of intercensal exposure times.\\ 
We considered the death count ($y_{tj}$) and the population exposed to risk ($n_{tj}$), for female and male separately, and with \textit{time} $\times$ \textit{age} intervals equal to 1-\textit{year} $\times$ 1-\textit{year}, for the following 37 countries: Australia, Austria, Belarus, Bulgaria, Canada, Chile, Czech Republic, Denmark, Estonia, Finland, France, Germany, Greece, Hungary, Iceland, Ireland, Israel, Italy, Japan, Latvia, Lithuania, Luxembourg, Netherlands, New Zealand, Norway, Poland, Portugal, Russia, Slovakia, Slovenia, Spain, Sweden, Switzerland, Taiwan, United Kingdom, Ukraine, and United States.  Belgium was omitted because of the presence of missing values in the data. The age domain of the data is the same for all the countries considered: 1-\textit{year} classes from $0$ to $110$, and a final cumulative class with individuals aged more than $110$ years. On the other hand, the time domain is not the same for each country: the Chilean data are the shortest series with the timespan of only $14$ years (1992-2005) while the Swedish dataset is the longest with $264$ years (1771-2014) of observations. The average timespan across all datasets is $88$ years. Table \ref{tab:tab_label} reports a summary.
\begin{center}
< Table \ref{tab:tab_label} >
\end{center}

\section{Results and Discussion}\label{s5}
\begin{figure}
\centering
\includegraphics[scale=0.5]{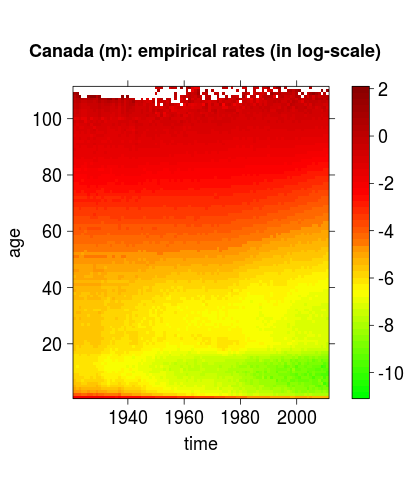} \includegraphics[scale=0.5]{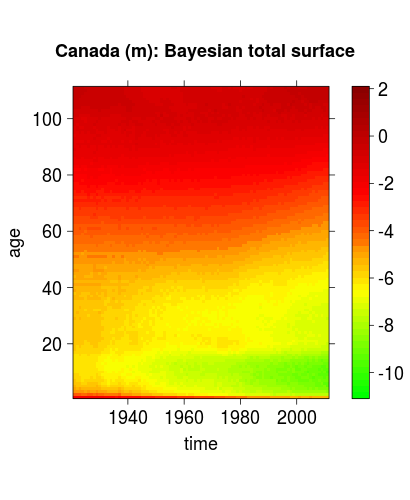}\\
\includegraphics[scale=0.5]{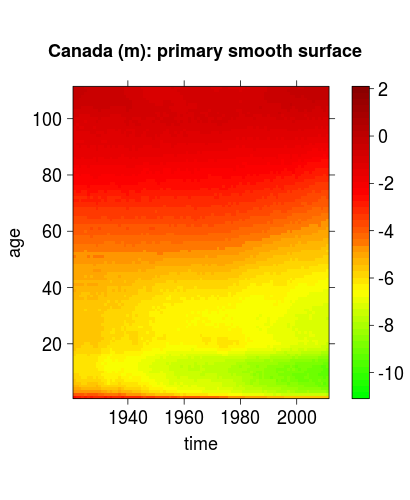} \includegraphics[scale=0.5] {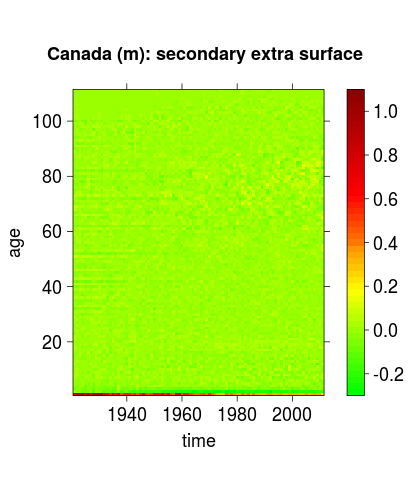}
\caption{Canada, male population. Empirical rates surface $\mathbf{s}_m$ (in log-scale, top left), Bayesian total surface $\mathbf{s}_b$ (top right), primary smooth surface $\mathbf{s}_1$ (bottom left), secondary extra surface $\mathbf{s}_2$ (bottom right).} \label{fig:canada}
\end{figure}

For each country, we estimated the two components $\mathbf{x}$ and $\mathbf{z}$ with the respective precisions $\gamma_x$ and $\gamma_z$, and produced three estimated mortality surfaces: the surface of the marginal posterior means $\mathbf{s}_b=\log(\hat{\mu}_0)+\mathbf{\hat{x}}+\mathbf{\hat{z}}$, the estimated primary smooth surface $\mathbf{s}_1=\log(\hat{\mu}_0)+\mathbf{\hat{x}}$, and the estimated secondary surface $\mathbf{s}_2=\mathbf{\hat{z}}$ representing additional mortality on top of $\mathbf{s}_1$. In addition, we considered also the surface of the empirical rates in log-scale $\mathbf{s}_m=\{\log(m_{tj}), (t,j) \in \mathcal{L} \}$ as comparison term. In Figure \ref{fig:canada} the case of Canada, male population, is presented as an illustration.
\begin{center}
< Figure. \ref{fig:canada} >
\end{center}
The Bayesian surface $\mathbf{s}_b$ (top right in Figure \ref{fig:canada}) is the overall estimate based on the raw empirical rates (in log-scale) $\mathbf{s}_m$ (top left in Figure \ref{fig:canada}). These two surfaces are rather similar in the case of Canada male and essentially in all our reconstructions, the purpose of this paper being the decomposition of $\mathbf{s}_b$ into $\mathbf{s}_1$ and $\mathbf{s}_2$, the smooth component and the additional free component. However, we see a difference between $\mathbf{s}_b$ and $\mathbf{s}_m$ in Figure \ref{fig:canada}, namely that in $\mathbf{s}_b$ estimates for the most elderly population are given, while empirical estimates are missing. When information is poor, particularly at the elderly ages with small death counts and small populations at risk, the empirical mortality rates are affected by large variations or can not be even calculated at the log-scale, so the white areas in the top border of the $\mathbf{s}_m$ surfaces. The presence in the Bayesian model of a spatial smooth component allows to estimate missing values as Bayesian averages of neighbouring rates.\\
The interesting contribution of our approach is the decomposition represented by $\mathbf{s}_1$ (bottom left in Figure \ref{fig:canada}) and $\mathbf{s}_2$ (bottom right in Figure \ref{fig:canada}). We see that $\mathbf{s}_1$ is a smooth surface in the Canadian male case, it is almost identical to $\mathbf{s}_b$, as $\mathbf{s}_2$ is in practice around zero everywhere. There is in this case a small exception, except for extra mortality at birth (first row in the bottom of $\mathbf{s}_2$). Because we have a constant smoothing strength in $\mathbf{s}_1$ (represented by a constant $\gamma_x$), adding this extra mortality to the smooth component $\mathbf{s}_1$ would apparently be problematic, as $\mathbf{s}_1$ would be less smooth between the first and second row, compared to the rest of the surface. Hence extra mortality in $\mathbf{s}_2$. 

\subsection{Sweden and Italy}
\begin{figure}
\centering
\includegraphics[scale=0.235]{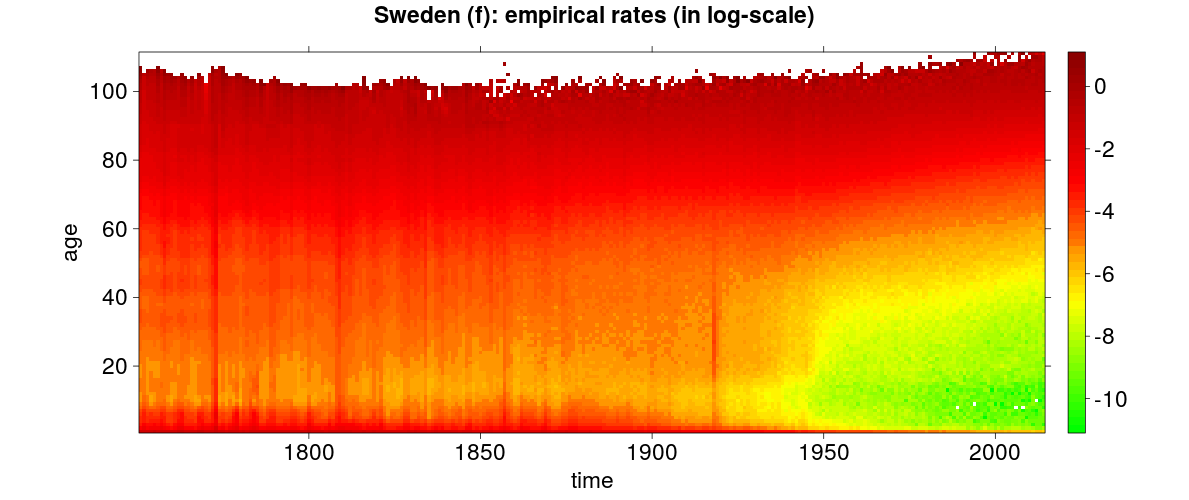} \includegraphics[scale=0.235]{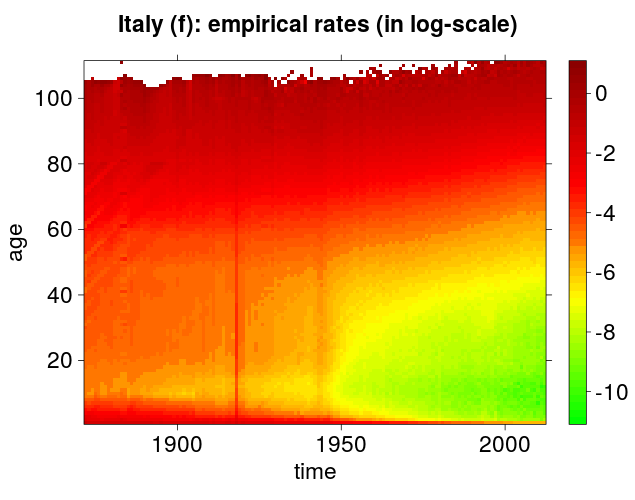}\\
\includegraphics[scale=0.235]{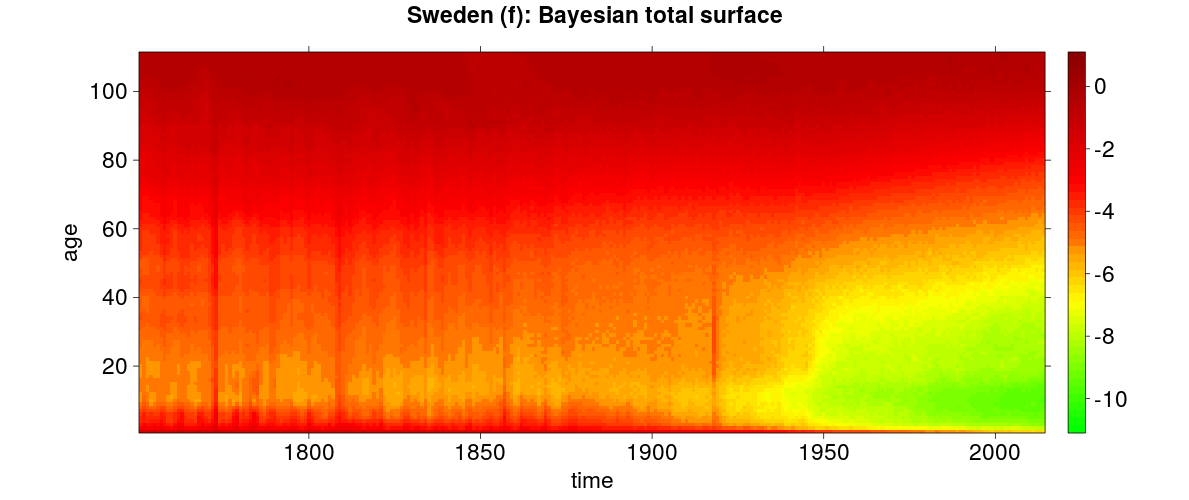} \includegraphics[scale=0.235]{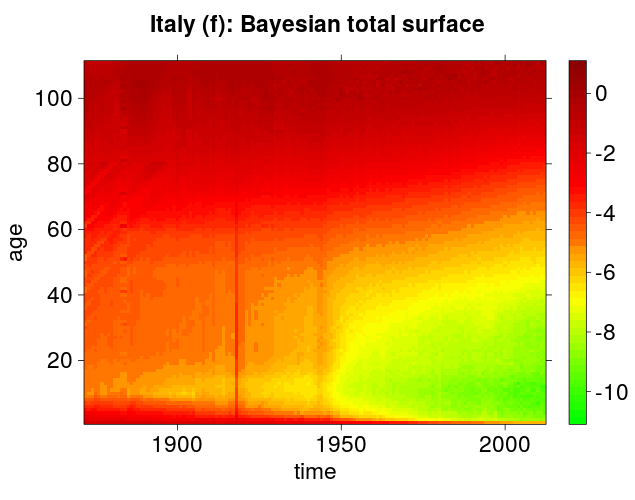}\\
\includegraphics[scale=0.235]{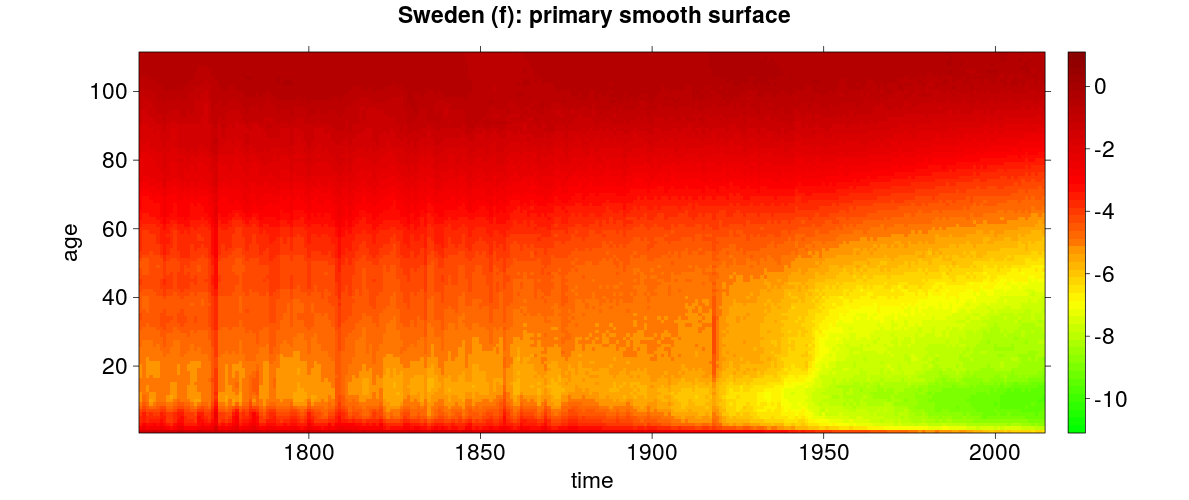} \includegraphics[scale=0.235]{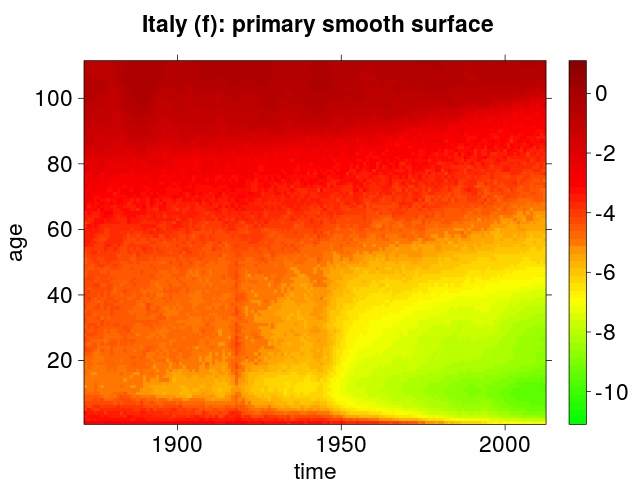}\\
\includegraphics[scale=0.235]{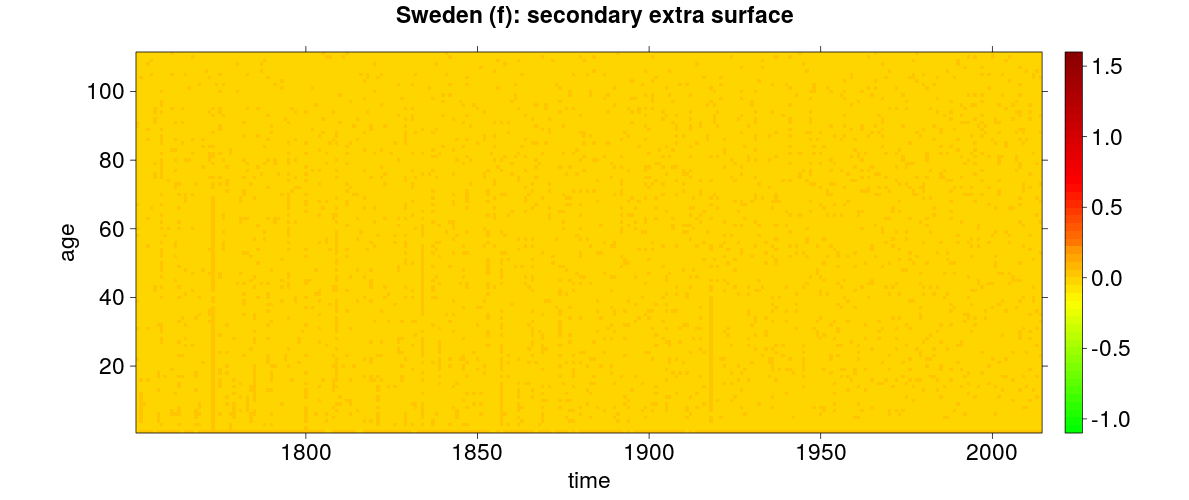} \includegraphics[scale=0.235] {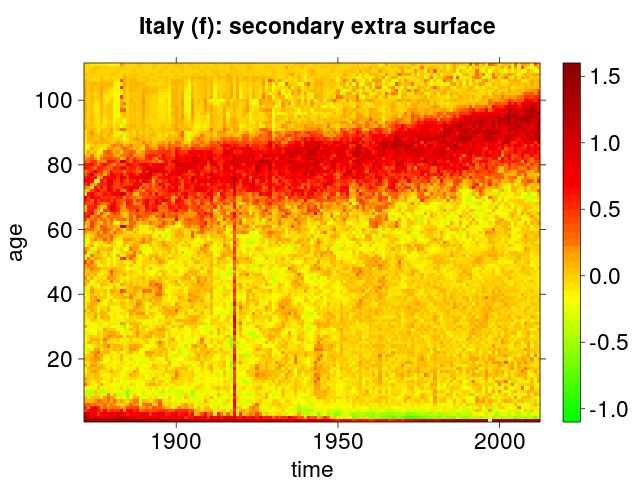}
\caption{Sweden (first column) and Italy (second column), female population. Empirical rates surface $\mathbf{s}_m$ (in log-scale, first row), Bayesian total surface $\mathbf{s}_b$ (second row), primary smooth surface $\mathbf{s}_1$ (third row), secondary extra surface $\mathbf{s}_2$ (fourth row).}\label{fig:female}
\end{figure}

\begin{figure}
\centering
\includegraphics[scale=0.235]{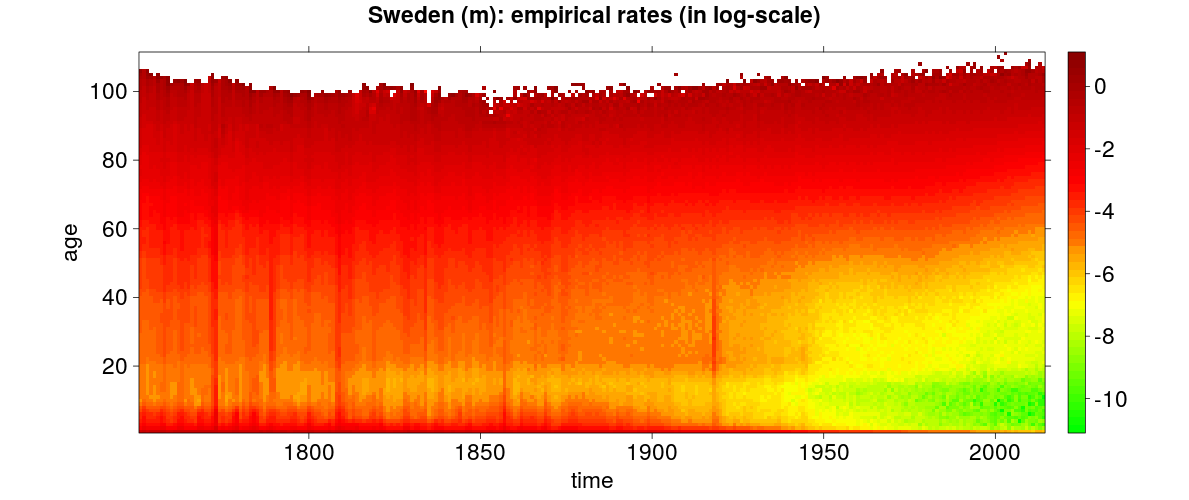} \includegraphics[scale=0.235]{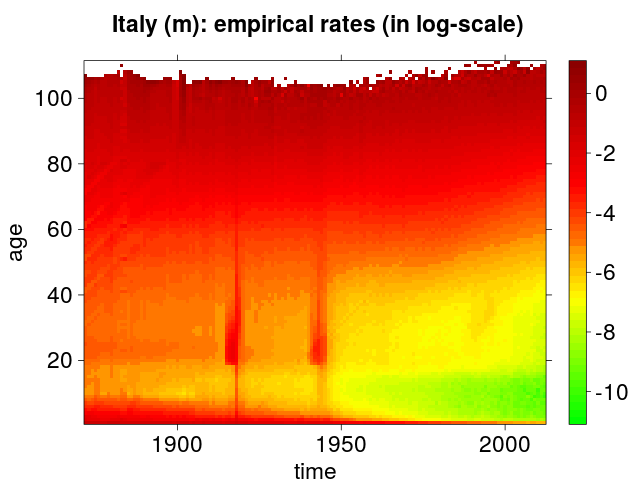}\\
\includegraphics[scale=0.235]{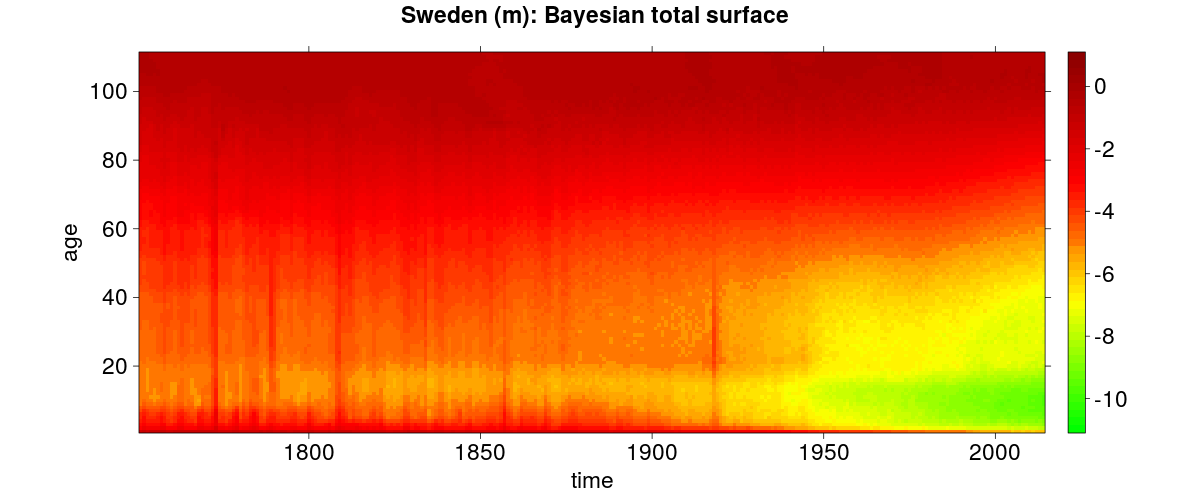} \includegraphics[scale=0.235]{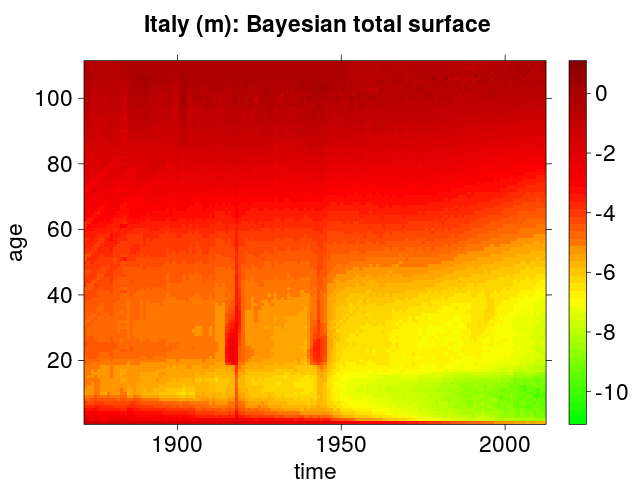}\\
\includegraphics[scale=0.235]{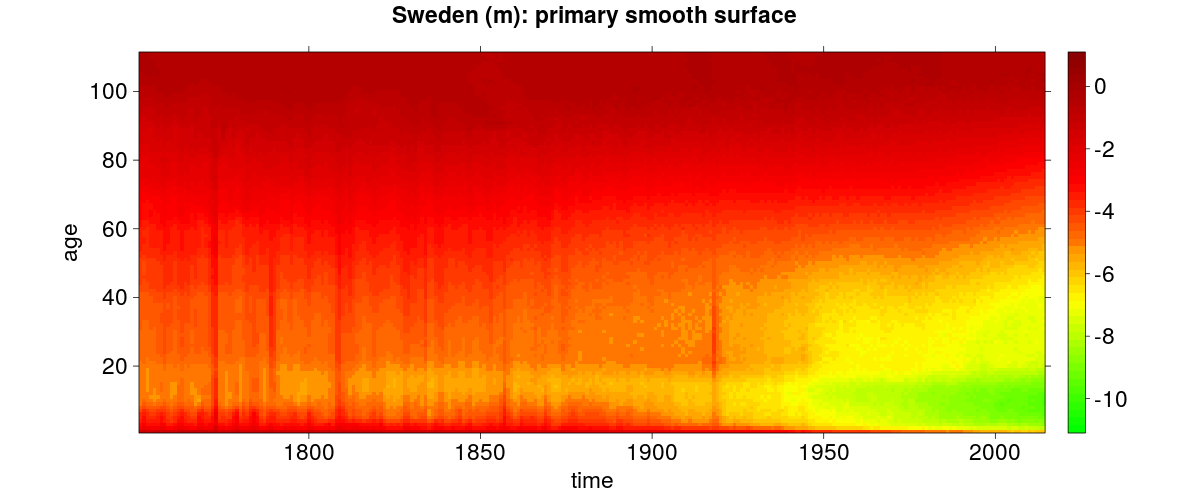} \includegraphics[scale=0.235]{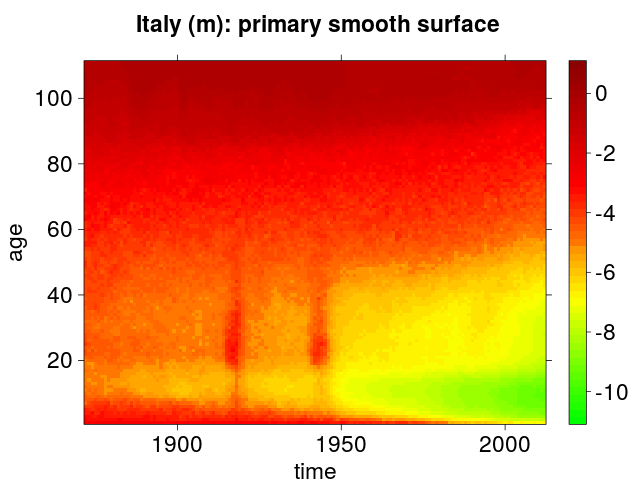}\\
\includegraphics[scale=0.235]{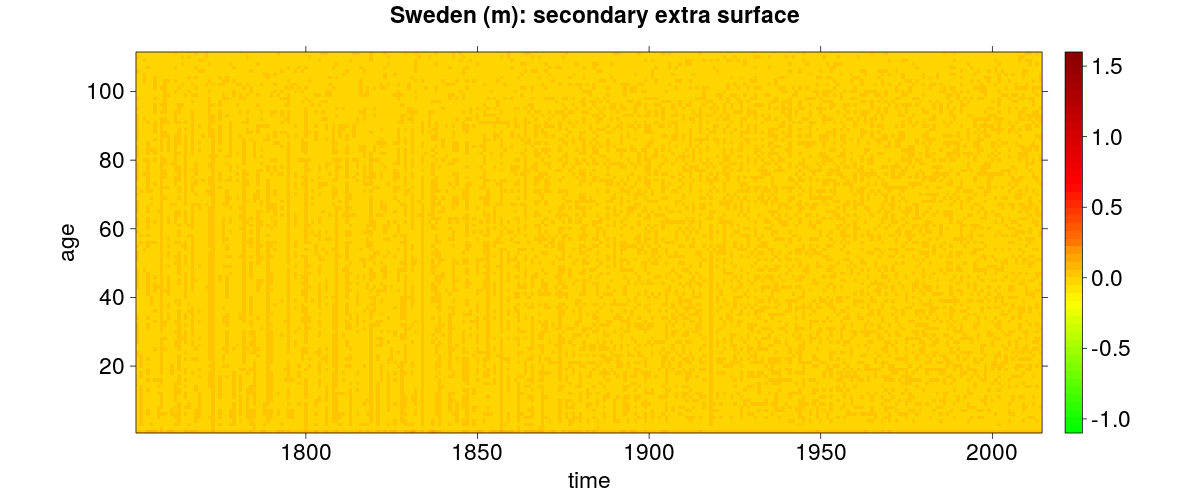} \includegraphics[scale=0.235]{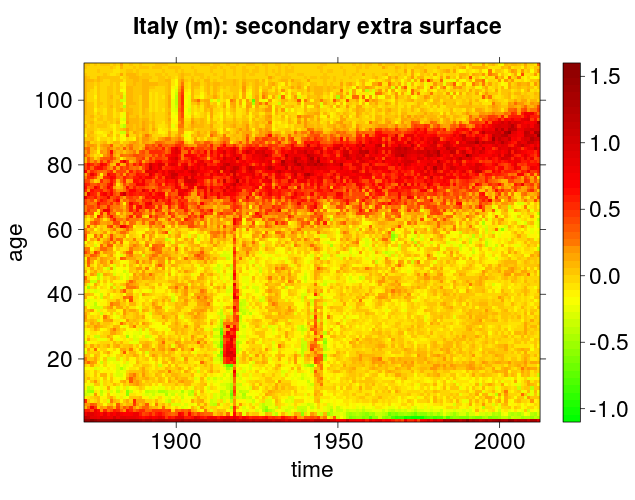}
\caption{Sweden (first column) and Italy (second column), male population. Empirical rates surface $\mathbf{s}_m$ (in log-scale, first row), Bayesian total surface $\mathbf{s}_b$ (second row), primary smooth surface $\mathbf{s}_1$ (third row), secondary extra surface $\mathbf{s}_2$ (fourth row).}\label{fig:male}
\end{figure}

Our analysis of all 37 countries led to two different typical patterns concerning the secondary surfaces and well represented by the examples of Sweden and Italy. In Figure \ref{fig:female} and Figure \ref{fig:male}, the surfaces of Sweden (first column) and Italy (second column), for female and male population respectively, are reported.
\begin{center}
< Figure \ref{fig:female} >
\end{center}
\begin{center}
< Figure \ref{fig:male} >
\end{center}
In the case of Sweden, the primary smooth surface $\mathbf{s}_1$ is the only important component and represents in detail the whole shape of the mortality pattern, as it was for Canada in Figure \ref{fig:canada}, corresponding to estimated values $\hat{z}_{tj}$ being close to zero (in the case of the male population some weak effect can be noticed). On the other hand, in the case of Italy, for both genders, the secondary component presents strong and interesting patterns. First, at the birth, particularly at the beginning of the period and across the whole time domain, the surface $\mathbf{s}_2$ shows relevant levels of additional mortality which are related to the infant mortality dynamics, and which appear to be stronger than what expected by an assumption of mortality smooth variation across the Lexis surface. Most importantly, we observe the presence of a large band across time, with a slightly positive slope and approximately in the age range 60-90. In this band the force of mortality accelerates with respect to the pattern explained by the primary component and displayed in the surface $\mathbf{s}_1$.\\
The visual analysis of mortality surfaces is also interesting to detect historical dynamics. For instance, in Figure \ref{fig:female} and Figure \ref{fig:male}, the effects of the 1918 Spanish Influenza epidemic (both countries) and of two world wars (Italy only) are clearly visible. Elevated mortality represented by thin vertical lines in 1918 can be seen in all graphs. In addition there is a clear effect for Italian men of World War I (1914-1918) and World War II (1940-1945). For Italian women the effect is very small. Sweden was neutral under both wars, and hence no war effects appear.
\begin{center}
< Table \ref{tab:tab_ratio} >
\end{center}

\subsection{The mortality profiles}
\begin{figure}
\centering
\includegraphics[scale=0.35]{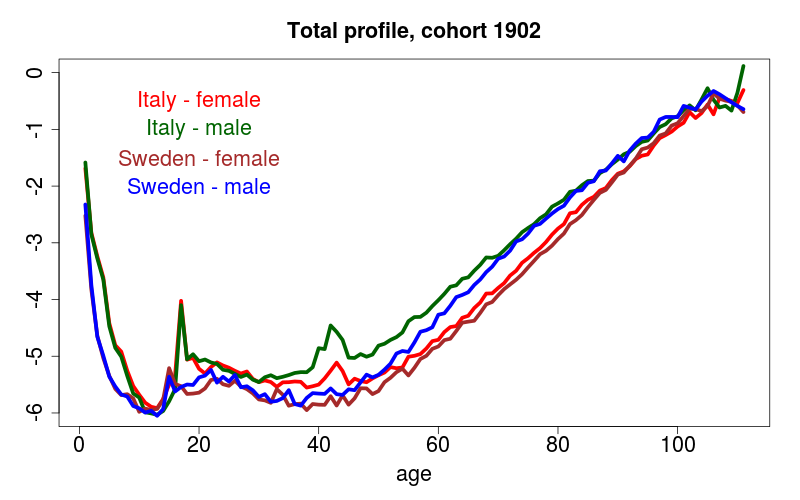}\\
\includegraphics[scale=0.35]{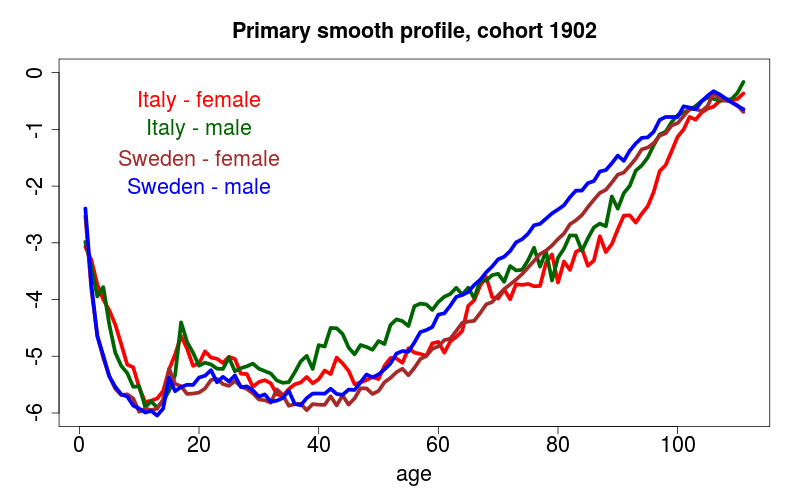}\\
\includegraphics[scale=0.35]{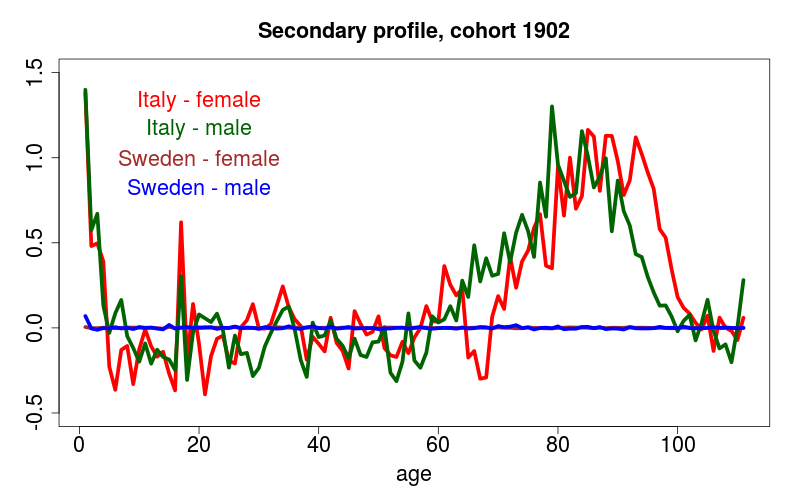}
\caption{Sweden and Italy, mortality profiles for the cohort ``1902'', levels in log-scale.}\label{fig:cohort}
\end{figure}

\begin{figure}
\centering
\includegraphics[scale=0.35]{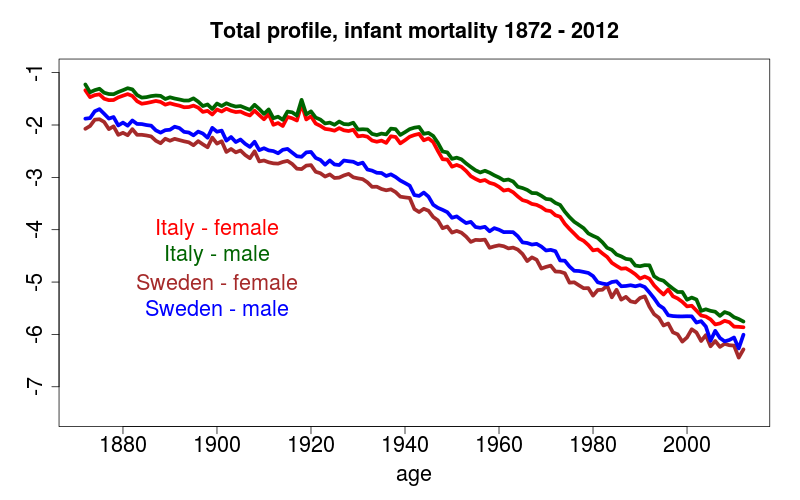}\\
\includegraphics[scale=0.35]{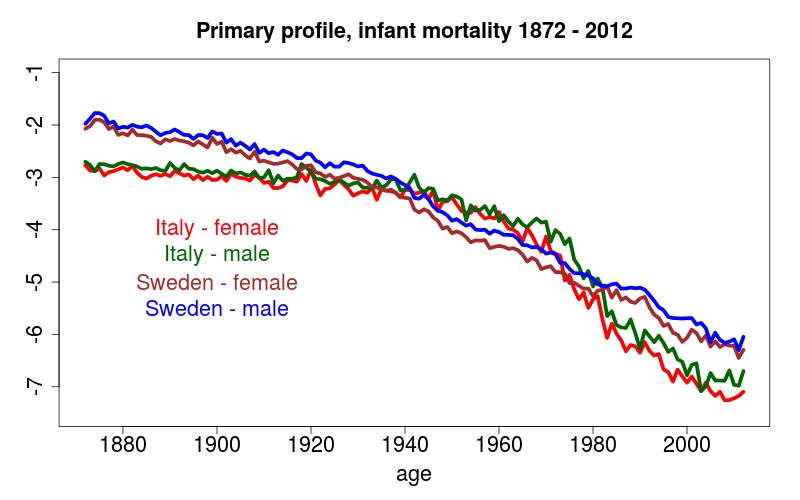}\\
\includegraphics[scale=0.35]{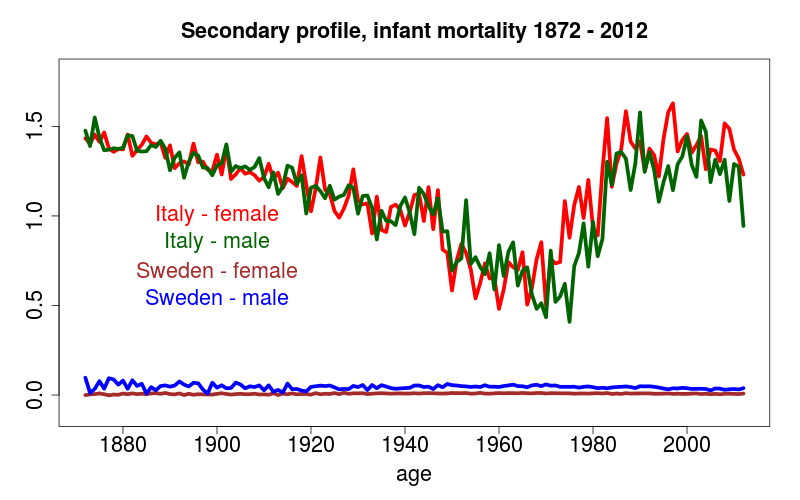}
\caption{Sweden and Italy, infant mortality profiles during the period 1872-2012, levels in log-scale.}\label{fig:infant}
\end{figure}

\begin{figure}
\centering
\includegraphics[scale=0.35]{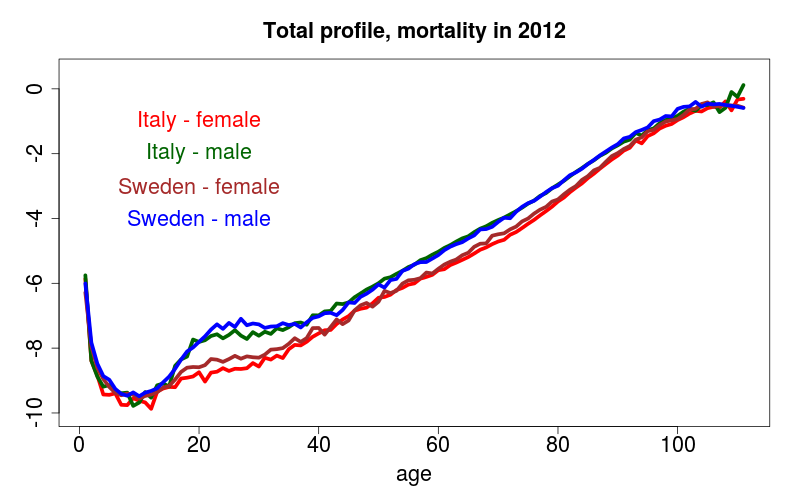}\\
\includegraphics[scale=0.35]{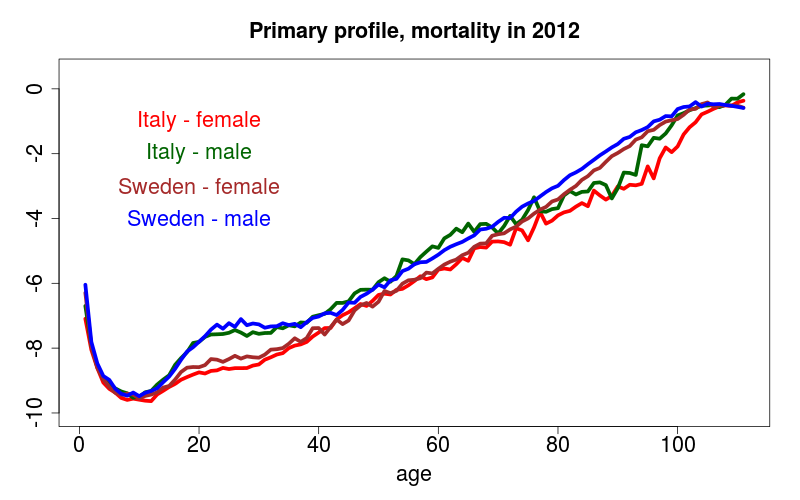}\\
\includegraphics[scale=0.35]{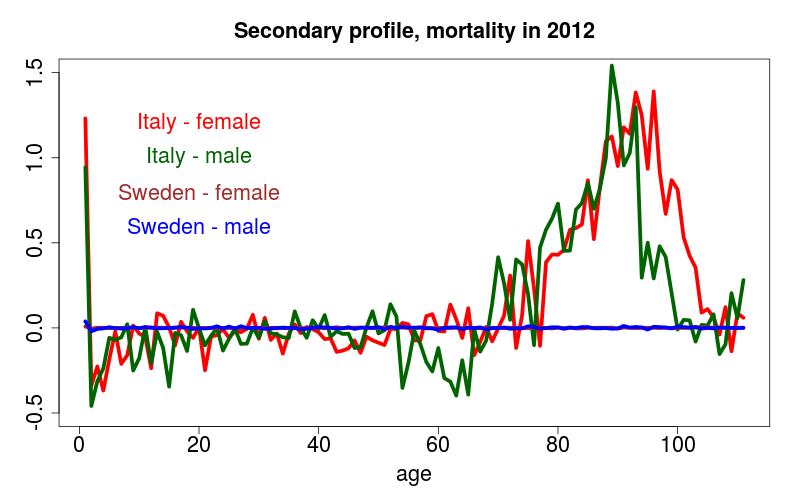}
\caption{Sweden and Italy, mortality profiles by age at the year 2012, levels in log-scale.}\label{fig:time}
\end{figure}

The Bayesian procedure allows also to decompose estimated mortality profiles for specific cohorts, for specific ages, or for specific time points. In Figure \ref{fig:cohort}, Figure \ref{fig:infant}, and Figure \ref{fig:time} we show, as examples, the mortality profiles of the cohort 1902, for the infant mortality during the period 1872-2012, and by age in the year 2012 respectively; for Sweden and Italy, female and male population.
\begin{center}
< Figure \ref{fig:cohort} >
\end{center}
\begin{center}
< Figure \ref{fig:infant} >
\end{center}
\begin{center}
< Figure \ref{fig:time} >
\end{center}
The significant presence of the extra mortality band is seen in Figure \ref{fig:cohort} and Figure \ref{fig:time} in the secondary profiles of Italy, compared to the null effect present in the plots of Sweden. In Figure \ref{fig:infant} we can observe the decreasing pattern of the infant mortality during the period 1872-2012, both in Italy and Sweden. It is interesting to notice the different patterns in the secondary profiles. In the case of Italy, female and male population, our decomposition allows to highlight a significant difference in the levels of mortality between the primary component and the secondary component: the infant mortality globally decreases (total and primary profiles) but since the 80s an increasing and then stable level of additional mortality with respect to the primary smooth profile is present in the Italian maps (secondary profiles). The extra mortality is in the range 0.5-1.5 in log-scale (corresponding to 1.6-4.5) compared to the baseline mortality rate in the magnitude of $-4$ in log-scale ($0.014$ and $0.015$ in mortality rate scale, for female and male respectively). This is an interesting observation that deserves further investigation.

\subsection{Computational details} 
\begin{figure}
\centering
\includegraphics[scale=0.27]{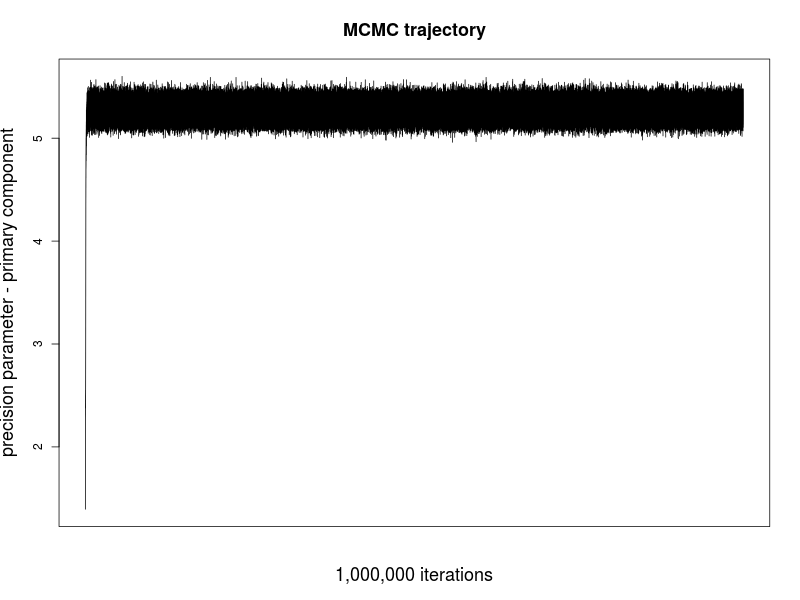}
\includegraphics[scale=0.27]{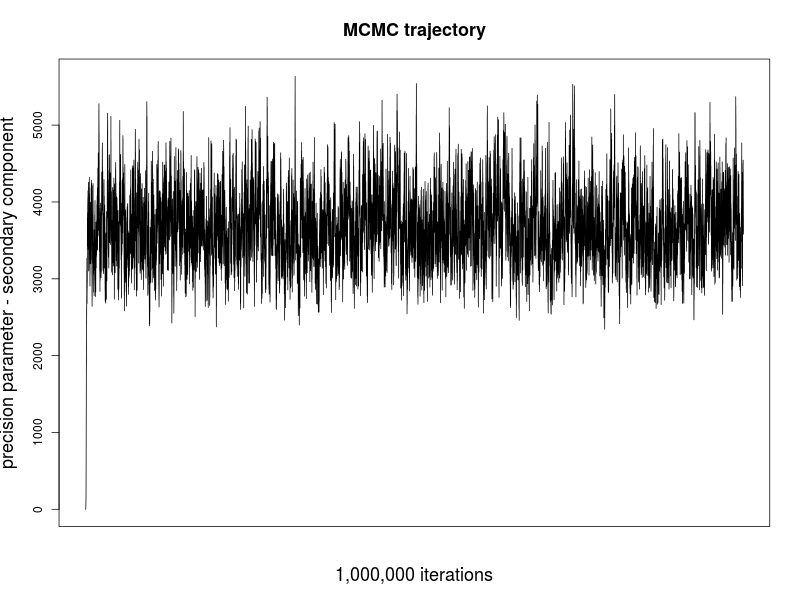}\\
\includegraphics[scale=0.27]{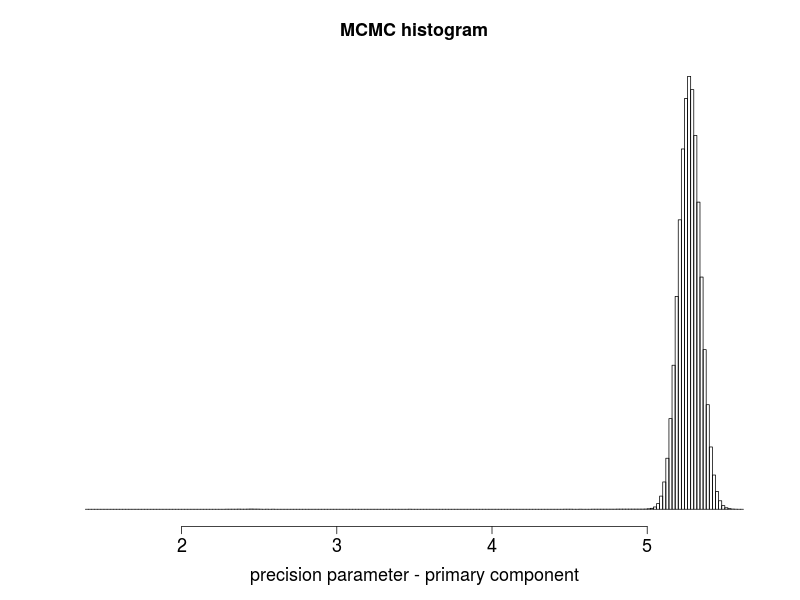}
\includegraphics[scale=0.27]{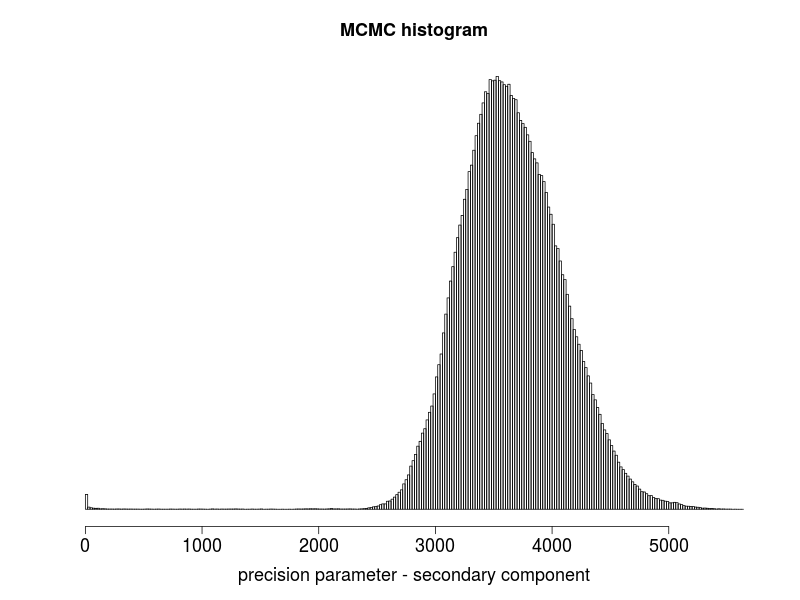}
\caption{MCMC convergence, simulation trajectories and posterior histograms of the precision parameters, Norway, male population, 1,000,000 iterations.}\label{fig:mcmc1}
\end{figure}

The convergence of the MCMC chains was inspected by visual checks and by the Gelman-Rubin statistics \citep{gelrub:1992}. In Figure \ref{fig:mcmc1} we show the MCMC trajectories and histograms of the two precision parameters in the case of the Norwegian data (male population), across 1,000,000 iterations. The second chain converges to a posterior marginal which is flatter than the posterior marginal of the first chain, but convergence is attained quickly for both parameters. In our applications, we used 100,000 total iterations: 70,000 of burn-in with the subsequent 30,000 MCMC samples used to compute the marginal posterior mean for each quantity of interest.
\begin{center}
< Figure \ref{fig:mcmc1} >
\end{center}
%

\subsection{The precision ratio}\label{parratio}
\begin{figure}
\centering
\includegraphics[scale=0.5]{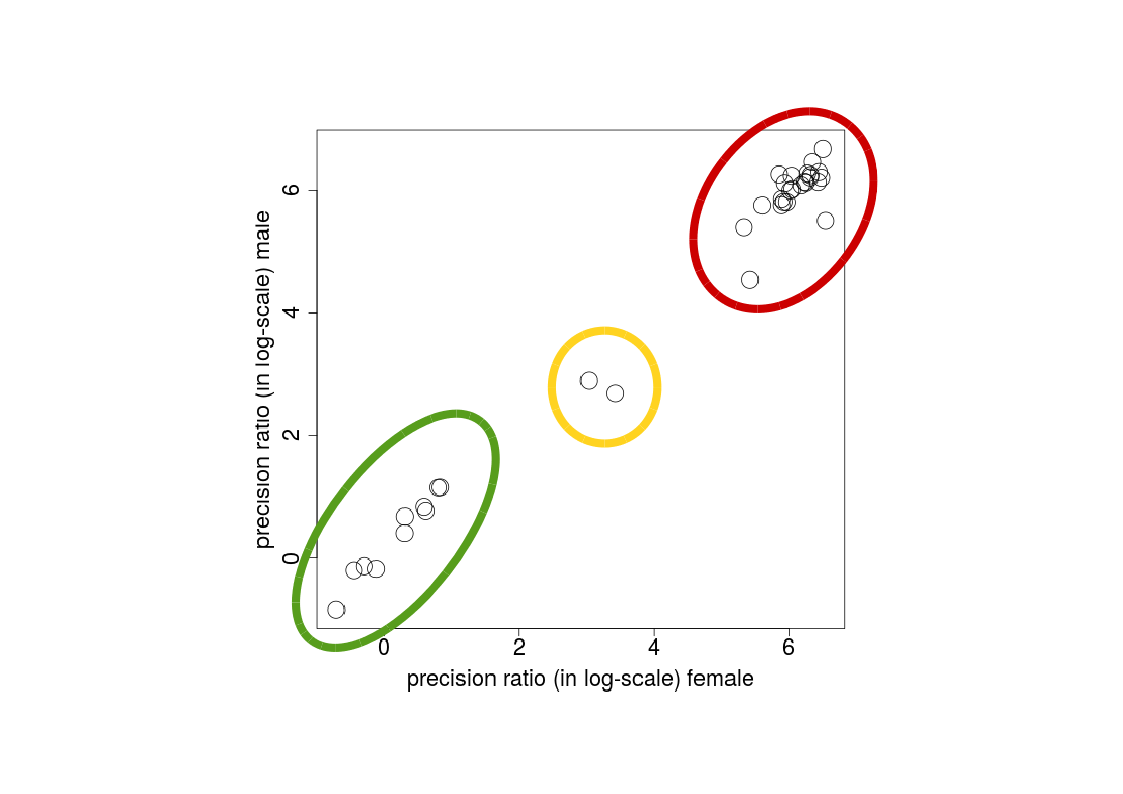}
\caption{precision ratio in log-scale, cross-plot by gender. The countries in the red cluster show no extra mortality and $\mathbf{s}_2$ is basically zero; for the countries in the green cluster we observed a band in $\mathbf{s}_2$; the two countries in the yellow cluster have no band, but some extra non smooth mortality. } \label{fig:ratio}
\end{figure}

The potential presence of extra mortality in the secondary surface can be detected also from the estimates of the two precision parameters $\gamma_x$ and $\gamma_z$. These parameters summarize the mortality variation across the whole Lexis graph. In fact, in each point $(t,j) \in \mathcal{L}$, the a priori conditional variance of $\log(\mu_{tj})$, given potential values $\mu_{si}$ in its neighbours $(s,i) \in \Delta_{tj}$, is equal to $\dfrac{\gamma_{x}^{-1}}{|\Delta_{tj}|}+\gamma_{z}^{-1}$  \citep{mollie:1999}, with $|\Delta_{tj}|$ denoting the number of knots in $\Delta_{tj}$. When the surface $\mathbf{s}_2$ is not relevant, the estimated values $\hat{z}_{tj}$ show small variation with a high value of the parameter $\gamma_z$. On the other hand, when in the secondary surface there is a relevant presence of additional mortality, the variation among the values $\hat{z}_{tj}$ increases, reducing the level of precision. As in \cite{mollie:1999}, we introduce a measure to assess the respective roles played by each component and define the following precision ratio 
$$\rho = \dfrac{\hat{\gamma}_z}{\hat{\gamma}_x},$$
that summarizes the information included in $\hat{\gamma}_x$ and $\hat{\gamma}_z$. In fact, small values of $\rho$ reflect the presence of non smooth variation and a secondary surface with relevant patterns whilst large values of $\rho$ denote that the variation of the primary smooth component dominates, with the surface $\mathbf{s}_2$ being not important.\\
In Table \ref{tab:tab_ratio} the estimates of the precision parameters and the precision ratio for each country, female and male population, are reported. There are ten countries (France, Germany, Italy, Japan, Poland, Russia, Spain, United Kingdom, Ukraine, and United States) with particularly small values of the precision ratio. They correspond to the cases in which we detected the presence of extra mortality in the secondary surface.\\
The information represented by the precision ratio can be also displayed graphically as in Figure \ref{fig:ratio} in which we plot the precision ratio at the log-scale and by gender. 
\begin{center}
< Figure \ref{fig:ratio} >
\end{center}
Three well defined clusters can be noticed. The first cluster (highlighted by a green circle) includes the set of countries mentioned above which present extra mortality in the surface $\mathbf{s}_2$, in the second cluster (yellow circle) there are Canada and Netherlands with a moderate presence of additional mortality, while all the other countries for which the mortality pattern is well explained by the primary component belong to the last cluster (red circle). In general, from the bottom left corner of the diagram and along the diagonal, the countries are displayed by increasing levels of the precision ratio, corresponding to decreasing levels of extra mortality.\\
Interestingly, the green cluster includes the most populated countries in our datasets, Canada and Netherlands with a mid size of population belong to the yellow cluster, while the countries in the red cluster have all small populations, see for example \cite{un:2015}. Therefore, it seems that there is a potential positive association between the presence of extra mortality in the secondary surface $\mathbf{s}_2$ and population size of the respective country.

\subsection{The overdispersion band}
\begin{figure}
	\centering
	\includegraphics[scale=0.35]{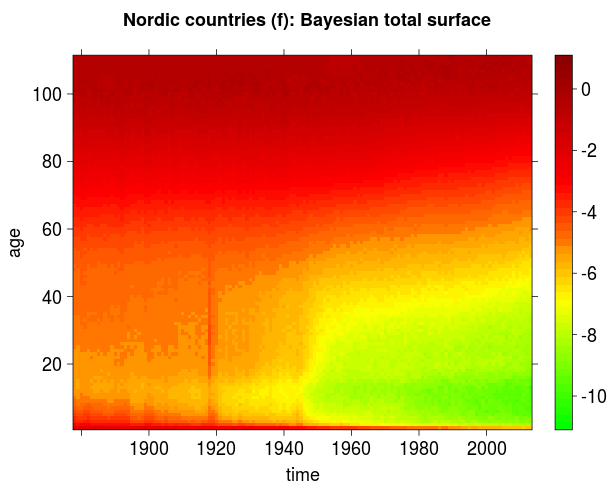} 	\includegraphics[scale=0.35]{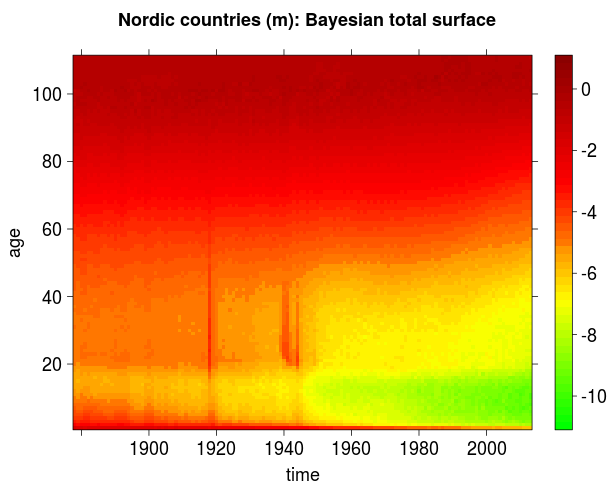}\\ 
	\includegraphics[scale=0.35]{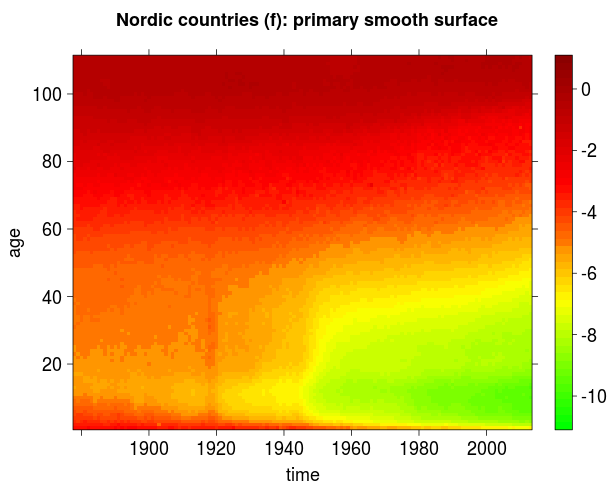}	\includegraphics[scale=0.35]{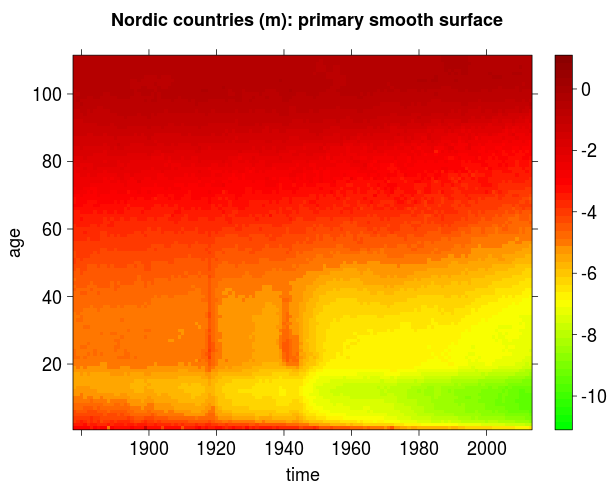}\\
	\includegraphics[scale=0.35]{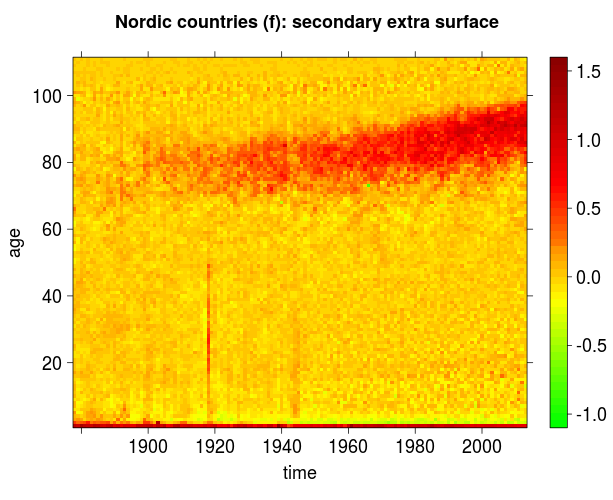} 	\includegraphics[scale=0.35]{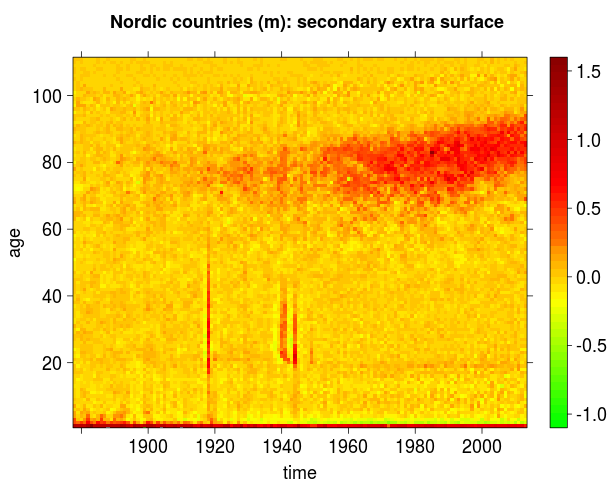}
	\caption{Nordic countries, female population (first column) and male population (second column). Bayesian total surface $\mathbf{s}_b$ (first row), primary smooth surface $\mathbf{s}_1$ (second row), secondary extra surface $\mathbf{s}_2$ (third row).} \label{fig:nordic}
\end{figure}

As we reported above, the countries in the green cluster in Figure \ref{fig:ratio} (France, Germany, Italy, Japan, Poland, Russia, Spain, United Kingdom, Ukraine, and United States) present in the secondary surface significant levels of extra mortality across the time domain, particularly at the birth and in a band with a slightly positive slope in the age interval 60-90.\\ 
In order to explain the relevance of these patterns, we need to further clarify the respective roles played by the two terms $\mathbf{x}$ and $\mathbf{z}$ in the estimation procedure. Both terms represent effects which are random in the Bayesian setting with appropriate priors: a smooth Gaussian Markov random field and a set of independent Gaussian variables respectively. Therefore, why would the estimation procedure need to assign additional mortality to some areas of the secondary surface? It is obvious that the use of the independent shocks $z_{tj}$, which by definition fit the data with a larger level of flexibility, would be necessary in addition to the smooth components $x_{tj}$ only when the marginal variation of the data in those areas is larger than in the rest of the surface. In particular, when this variation is not compatible with the smoothing level of the rest of the surface. Hence, the mortality patterns potentially observable in the secondary surface represent patterns of overdispersion \citep{cox:1983, xekala:2014}, beyond what the Gaussian Markov random field can accommodate. Therefore,  $\mathbf{s}_2$ is not a residual surface but it accounts for additional mortality patterns with a level of variation larger than the level in $\mathbf{s}_1$.\\
Overdispersion, especially with count data, is a well known phenomenon \citep{breslo:1984, lindse:1995}. It arises when the data exhibit a larger variation than expected under a reference model and may be generally determined by several factors: latent dependence among the elements of the population, contagion and change of the individual behaviours during the survey, clustering in the structure of the population and heterogeneity inside the population \citep{xekala:2014}. In the particular case of spatial count data, the overdispersion may be determined by two relevant causes as pointed out in \cite{clayal:1993}: the clustering and the heterogeneity acting on the response counts across the spatial domain of interest, that in our case is the Lexis graph. Therefore, following \cite{clayal:1993}, we can interpret the model in Equation \eqref{log3} as a log-linear model with constant average $\mu_0$ and with two random terms, $x_{tj}$ and $z_{tj}$, accounting for extra variation due to the clustering and extra variation due to the heterogeneity respectively. Hence, while the primary surface $\mathbf{s}_1$ shows the smooth pattern in terms of overdispersion effects by clustering, the patterns potentially present in the secondary surface $\mathbf{s}_2$ represent overdispersion effects due to the heterogeneity in the counts and display specific features which are locally non smooth as the infant mortality dynamics or the excess of mortality during the period of the Spanish Influenza epidemic for instance.\\
In this setting the overdispersion band in the age interval 60-90 can be interpreted as a heterogeneity effect due to large counts of deaths. In fact, we notice that the bands occur in large countries in ages 60-80 to begin with, and in ages 70-90 in recent years. These are the age intervals in which many individuals die (see also the plots in Figure \ref{fig:cohort} and Figure \ref{fig:time}), with higher marginal variation and because of many different causes of death.\\
Furthermore, the slope agrees with an increasing mean age at death (i.e. observed deaths). Over the years, the bands show a tendency to become more concentrated around the mean age. This is in agreement with what we know about the age distribution of mortality in many countries: not only an increase in life expectancy/mean age at death, but also a compression of mortality around the mean.\\ 
The bands are not visible in countries with small populations because overdispersion is difficult to detect when the counts are small as pointed ou by \cite{daviso:2008}: ``large amounts of data will be needed to detect overdispersion when the counts are small''. Indeed, when data of small countries are aggregated with respect to a shared time domain, as we did with Nordic countries (Denmark, Finland, Iceland, Norway, and Sweden) across the period 1878-2013 for both female and male populations, then significant levels of extra mortality become clearly visible in the secondary surface, as displayed in Figure \ref{fig:nordic}. Each country alone does not present a relevant secondary pattern $\mathbf{z}$ but when the data are aggregated, the increasing in size makes the overdispersion band visible. Therefore, the bands do not appear in countries with small populations, because potential overdispersion is difficult to detect when the population exposed to risk is small.

\begin{center}
	< Figure \ref{fig:nordic} >
\end{center}

\subsection{The Vaupel model}
As mentioned in Section \ref{s2}, the idea to decompose the force of mortality into two components was introduced by \cite{vaupel:1979}. Indeed, when represented in log-scale, the model in Equation \eqref{vaupel} can be rewritten as
$$
\log(\mu_{tj})=\log(\mu_{tj,1})+\log(\zeta_{tj}),
$$
which is very similar with Equation \eqref{log3}. In particular, $\mu_{tj,1}$ is the force of mortality of a standard individual with unit frailty and, for varying $t$ and $j$, represents the main standard pattern of mortality across the Lexis graph, as the primary surface $\mathbf{s}_1$ previously introduced. Therefore, in each point of the Lexis graph the term $\log(\mu_{tj,1})$ and the quantity $\log(\mu_0)+x_{tj}$ play similar roles as they account for the main trend over time and age.\\
The second term in the Vaupel model $\zeta_{tj}$ represents the average frailty at the population level for the surviving individuals in the cohort $c=t-j$. \cite{vaupel:1979} consider that each individual in the cohort $c$, at any point $(t,j)$ of the Lexis graph, has frailty $u$ which is a random quantity and Gamma distributed with parameters $(k_{tj},\nu_{tj})$. These parameters are fixed to initial values $(k_{t0},\nu_{t0})$ at the time of birth and then updated by the recursive use of the cumulative hazard function, see \cite{vaupel:1979} for details (in the notation of their paper the parameters are considered depending only on the age $j$ as they are the same for every cohort $c$ and independent of the time point $t$). Under this assumption the term $\zeta_{tj}$ is derived as the expectation
$$
M_{tj}[u]=\dfrac{k_{tj}}{\nu_{tj}}
$$
where $M_{tj}$ denotes the operator of expectation at the Lexis point $(t,j)$ with the respective Gamma distribution.
As \cite{vaupel:1979} assert, ``the Gamma distribution was chosen because it is analytically tractable and readily computable'', but its shape is not very different from the Log-Normal distribution that was also considered by the authors as a potential choice to model the individual frailty. Therefore the term $\zeta_{tj}$ and the secondary component $z_{tj}$ (at the exponential scale) originate from similar distributions, Gamma vs. Log-Normal, and play similar roles as they both account for additional mortality and ``acknowledging the known heterogeneity in populations'' \citep{vaupel:1979}.\\
Thus, if we consider the following similitude 
$$
\log(\mu_{tj,1}) \sim \log(\mu_{0})+x_{tj},
$$
and
$$
\log(\zeta_{tj}) \sim z_{tj},
$$
in terms of roles played in the respective frameworks, it is clear that our model is a simple extension of the Vaupel model in the direction of assuming that the term $\log(\mu_{tj,1})$ is a Markov random field smooth across the Lexis graph. Therefore, we also see that the secondary component $z_{tj}$ can be interpreted as frailty at the population level (in log-scale).

\section{Conclusion}\label{s6}
The analysis of mortality data in terms of surface over a Lexis structure allows to capture the simultaneous effects of period, age, and cohort. Following an approach originally developed in spatial statistics, we proposed a model which allows to decompose the force of mortality into two components: a primary smooth component that gives the main features of age-specific mortality trends over time and a secondary component without any Lexis structure that represents additional mortality patterns not captured by the main trend.\\
The model was estimated in a Bayesian framework. Therefore the two components were specified in terms of prior distributions: a set $\mathbf{x}$ of conditional autoregressive terms jointly distributed as an intrinsic Gaussian Markov random field and a set $\mathbf{z}$ of independent Gaussian variates respectively. The smoothing of the primary component and the flexibility of the secondary component were controlled by the respective precisions $\gamma_x$ and $\gamma_z$, included into the model as hyperparameters with vague Gamma priors. In order to compute Bayesian estimates we used MCMC computation with Metropolis and Gibbs sampler steps.\\ 
It is important to emphasize that a Bayesian approach can be sensitive to the choice of priors and hyperpriors. In our model, a relevant question concerns the distribution of the precision parameters $\gamma_x$ and $\gamma_z$. Indeed, they modulate the trade-off between smoothing and no smoothing across the whole Lexis surface and, although they are sensitive to the choice of the respective hyperparameters, the effect is not on the estimation of the force of mortality but mainly on the variance of this estimation \citep{pental:2003}. Thus, the mortality surfaces based on the Bayesian posterior estimates are at most weakly affected by prior choices on the hyperparameters.\\
We presented an extensive application to data from the Human Mortality Database, where 37 countries are considered. For each country, three mortality surfaces were computed: the Bayesian surface, estimating the whole force of mortality; the primary surface, smooth across period, age, and cohort; and the secondary surface of potential additional mortality. We interpreted the potential extra mortality in the secondary surface as an overdispersion effect. A significant amount of extra mortality in the secondary component is necessary only when the smooth component alone is not able to account for the variation of the data.\\ 
Particularly interesting are two relevant patterns over the time domain: an excess of mortality at the birth related to the infant mortality dynamics and a band of overdispersion in the age interval 60-90 with a slightly positive slope in time in agreement with an increasing mean age at death and a compression of mortality around the mean. The patterns were detectable only in countries with large population size. In countries with small population size those patterns are not absent, they are simply hidden and the lack of detection is due to the small amount of statistical information in terms of small counts.\\
There can be several reasons behind the excess mortality represented by the bands: of course one possible hypothesis is that better health systems delay the death of a part of the elderly population, which therefore has to die ``in excess'' in the later years. The slight positive trend of the band might indicate that these excess deaths are further delayed. We notice that these effects are in addition to the general smooth trends in population aging, which are present in the primary smooth component. \\ 
The issue is also related to the similitude between our Bayesian decomposition and the model proposed by \cite{vaupel:1979}, similitude that allows to interpret the secondary surface in terms of frailty patterns at the population level. These relevant questions deserve further investigations.

\clearpage

\bibliographystyle{natbib}

\newpage
\begin{table}
\footnotesize
\centering
\caption{Human Mortality Database, timespan for each country.}\label{tab:tab_label}
\begin{tabular}{lcr|lcr}
\toprule
Country & Time interval & N. years & Country & Time interval & N. years  \tabularnewline
\midrule
Australia   & 1921-2014 &  91 & Austria     & 1947-2014 &  68 \tabularnewline
Belarus	    & 1959-2014 &  56 & Bulgaria    & 1947-2010 &  64 \tabularnewline
Canada	    & 1921-2011 &  91 & Chile	    & 1992-2005 &  14 \tabularnewline
Czech Rep.  & 1950-2014 &  65 & Denmark     & 1835-2014 & 180 \tabularnewline
Estonia	    & 1959-2013 &  55 & Finland     & 1878-2014 & 137 \tabularnewline
France	    & 1816-2014 & 199 & Germany     & 1990-2013 &  23 \tabularnewline
Greece	    & 1981-2013 &  33 & Hungary     & 1950-2014 &  65 \tabularnewline
Iceland	    & 1838-2013 & 176 & Israel      & 1983-2014 &  32 \tabularnewline
Ireland     & 1950-2014 &  65 & Italy       & 1872-2012 & 141 \tabularnewline
Japan       & 1947-2014 &  66 & Latvia	    & 1959-2013 &  55 \tabularnewline
Lithuania   & 1959-2013 &  55 & Luxembourg  & 1960-2014 &  55 \tabularnewline
Netherlands & 1850-2012 & 163 & New Zealand & 1948-2013 &  66 \tabularnewline
Norway      & 1846-2014 & 169 & Poland      & 1958-2014 &  57 \tabularnewline
Portugal    & 1940-2012 &  73 & Russia	    & 1959-2014 &  56 \tabularnewline
Slovakia    & 1950-2014 &  65 & Slovenia    & 1983-2014 &  32 \tabularnewline
Spain	    & 1908-2014 & 107 & Sweden	    & 1751-2014 & 264 \tabularnewline
Switzerland & 1876-2014 & 139 & Taiwan	    & 1970-2014 &  41 \tabularnewline
United Kingdom & 1922-2013 &  92 & Ukraine	    & 1959-2013 &  65 \tabularnewline
United States  & 1933-2014 &  82 &              &           &     \tabularnewline
\bottomrule
\end{tabular}
\end{table}

\newpage
\begin{table}
\footnotesize
\centering
\caption{Precision parameter estimates and precision ratio for each country, female and male populations.}\label{tab:tab_ratio}
\begin{tabular}{lrrr|rrr}
\toprule
&                &Female          &      &                &Male            &       \tabularnewline
\midrule
Country    &$\hat{\gamma}_x$&$\hat{\gamma}_z$&$\rho$&$\hat{\gamma}_x$&$\hat{\gamma}_z$& $\rho$\tabularnewline
\midrule
Australia  &    3.9 & 2158.0 &	555.2 &	 3.9 &	1995.0 &	517.5 \tabularnewline
Austria	   &    3.3	& 1728.1 &	529.8 &	 3.2 &	1710.8 &	530.6 \tabularnewline
Belarus	   &	3.7	& 1782.2 &	482.9 &	 4.0 &	1740.1 &	438.3 \tabularnewline
Bulgaria   &	3.6	& 1808.3 &	508.8 &	 3.7 &	1705.0 &	462.2 \tabularnewline
Canada	   & 	4.4	&   90.6 &	 20.8 &	 4.3 &    78.0 &     18.1 \tabularnewline
Chile	   &	3.5	&  724.4 &	206.6 &	 3.7 &	 821.2 &	220.6 \tabularnewline
Czech Rep. &	3.3	& 1401.5 &	419.5 &	 3.4 &	1725.5 &	511.4 \tabularnewline
Denmark	   &	4.2	& 2815.0 &	667.2 &	 3.9 &	3069.4 &	794.6 \tabularnewline
Estonia	   & 	3.5	& 1421.1 &	411.1 &	 3.6 &	1440.3 &	402.9 \tabularnewline
Finland	   &	4.5	& 2822.6 &	621.0 &	 3.8 &	1763.3 &	461.2 \tabularnewline
France	   &	3.8	&    6.9 &	  1.8 &	 3.4 &	   7.8 &	  2.3 \tabularnewline
Germany	   &	3.7	&    2.8 &	  0.7 &	 3.3 &	   2.9 &	  0.9 \tabularnewline
Greece	   &	3.2	& 1169.0 &	361.6 &	 3.4 &	1199.0 &	349.5 \tabularnewline
Hungary	   &	2.9	& 1241.6 &	422.7 &	 2.9 &	1201.6 &	411.0 \tabularnewline
Iceland	   &	1.8	& 1142.8 &	630.7 &	 1.9 &	1014.1 &	543.5 \tabularnewline
Israel	   &	3.1	& 1157.2 &	372.8 &	 3.3 &	1105.4 &	333.8 \tabularnewline
Ireland    &    2.5 & 1122.8 &  443.0 &  2.7 &  1645.7 &   607.1  \tabularnewline
Italy	   &	3.4	&    6.3 &	  1.9 &	 3.1 &	   6.6 &	  2.2 \tabularnewline
Japan	   &	3.0	&    2.7 &	  0.9 &	 3.2 &	   2.7 &	  0.8 \tabularnewline
Latvia	   &	4.0	& 1377.7 &	348.4 &	 4.1 &	2142.3 &	522.9 \tabularnewline
Lithuania  &	3.6	& 1342.3 &	378.0 &	 3.8 &	1698.0 &	452.3 \tabularnewline
Luxembourg &	2.8	&  998.6 &	360.9 &	 2.8 &	 872.6 &	316.7 \tabularnewline
Netherlands&	4.9	&  151.1 &	 30.7 &	 4.6 &	  67.4 &	 14.7 \tabularnewline
New Zealand&	3.1	& 1607.1 &	518.4 &	 3.1 &	1427.6 &	456.5 \tabularnewline
Norway	   &	5.5	& 3152.9 &	570.1 &	 5.3 &	3393.6 &	642.5 \tabularnewline
Poland     &	2.4	&    5.4 &	  2.2 &	 2.8 &	   9.0 &	  3.2 \tabularnewline
Portugal   &	4.0	&  911.9 &	225.9 &	 4.4 &	 416.1 &	 93.8 \tabularnewline
Russia	   &	3.7	&    2.3 &	  0.6 &	 3.7 &	   3.0 &	  0.8 \tabularnewline
Slovakia   &	2.9	& 1608.9 &	552.5 &	 3.1 &	1552.9 &	498.2 \tabularnewline
Slovenia   &	3.9	& 1065.9 &	270.9 &	 4.2 &	1311.6 &	314.8 \tabularnewline
Spain	   &	3.4	&    7.9 &	  2.3 &	 3.6 &	  11.6 &	  3.2 \tabularnewline
Sweden	   &	4.7	& 3271.1 &  696.0 &	 4.4 &	1082.8 &    246.1 \tabularnewline
Switzerland&	3.9	& 2574.6 &	653.8 &	 3.7 &	1828.8 &	494.9 \tabularnewline
Taiwan	   &	5.8	& 2272.6 &	390.5 &	 6.1 &	2034.4 &	332.1 \tabularnewline
United Kingdom &	2.8	&    3.8 &	  1.4 &	 2.7 &	   4.0 &	  1.5 \tabularnewline
Ukraine	       &	3.3	&    4.4 &	  1.4 &	 3.4 &	   6.6 &	  2.0 \tabularnewline
United States  &	3.6	&    1.8 &	  0.5 &	 4.0 &	   1.7 &	  0.4 \tabularnewline
\bottomrule
\end{tabular}
\end{table}

\clearpage

\newpage

\newpage

\newpage

\newpage

\newpage

\newpage

\newpage

\newpage

\newpage


\newpage

\end{document}